%% file: mpENC.tex
\def\sphinxdocclass{report}
\documentclass[a4paper,10pt,openany,oneside]{sphinxmanual}
\usepackage{iftex}

\ifPDFTeX
  \usepackage[utf8]{inputenc}
\fi
\ifdefined\DeclareUnicodeCharacter
  \DeclareUnicodeCharacter{00A0}{\nobreakspace}
\fi
\usepackage{cmap}
\usepackage[T1]{fontenc}
\usepackage{amsmath,amssymb,amstext}
\usepackage[english]{babel}
\usepackage{times}
\usepackage[Bjarne]{fncychap}
\usepackage{longtable}
\usepackage{sphinx}
\usepackage{multirow}
\usepackage{eqparbox}

\addto\captionsenglish{}
\addto\captionsenglish{}
\SetupFloatingEnvironment{literal-block}{name=Listing }

\addto\extrasenglish{}

\input{unicode.tex}
\usepackage{enumitem}
\setlist[description]{style=nextline}

\title{Multi-Party Encrypted Messaging Protocol design document}
\date{19 January  2016}
\release{0.1-4-g75e6b2d}
\author{Mega Limited, Auckland, New Zealand \and Ximin Luo <xl@mega.co.nz> \and Guy Kloss <gk@mega.co.nz>}
\newcommand{\sphinxlogo}{}
\renewcommand{\releasename}{Release}
\makeindex

\makeatletter
\def\PYG@reset{\let\PYG@it=\relax \let\PYG@bf=\relax%
    \let\PYG@ul=\relax \let\PYG@tc=\relax%
    \let\PYG@bc=\relax \let\PYG@ff=\relax}
\def\PYG@tok#1{\csname PYG@tok@#1\endcsname}
\def\PYG@toks#1+{\ifx\relax#1\empty\else%
    \PYG@tok{#1}\expandafter\PYG@toks\fi}
\def\PYG@do#1{\PYG@bc{\PYG@tc{\PYG@ul{%
    \PYG@it{\PYG@bf{\PYG@ff{#1}}}}}}}
\def\PYG#1#2{\PYG@reset\PYG@toks#1+\relax+\PYG@do{#2}}

\expandafter\def\csname PYG@tok@nf\endcsname{\def\PYG@tc##1{\textcolor[rgb]{0.02,0.16,0.49}{##1}}}
\expandafter\def\csname PYG@tok@kd\endcsname{\let\PYG@bf=\textbf\def\PYG@tc##1{\textcolor[rgb]{0.00,0.44,0.13}{##1}}}
\expandafter\def\csname PYG@tok@bp\endcsname{\def\PYG@tc##1{\textcolor[rgb]{0.00,0.44,0.13}{##1}}}
\expandafter\def\csname PYG@tok@nc\endcsname{\let\PYG@bf=\textbf\def\PYG@tc##1{\textcolor[rgb]{0.05,0.52,0.71}{##1}}}
\expandafter\def\csname PYG@tok@sd\endcsname{\let\PYG@it=\textit\def\PYG@tc##1{\textcolor[rgb]{0.25,0.44,0.63}{##1}}}
\expandafter\def\csname PYG@tok@nv\endcsname{\def\PYG@tc##1{\textcolor[rgb]{0.73,0.38,0.84}{##1}}}
\expandafter\def\csname PYG@tok@w\endcsname{\def\PYG@tc##1{\textcolor[rgb]{0.73,0.73,0.73}{##1}}}
\expandafter\def\csname PYG@tok@c1\endcsname{\let\PYG@it=\textit\def\PYG@tc##1{\textcolor[rgb]{0.25,0.50,0.56}{##1}}}
\expandafter\def\csname PYG@tok@ch\endcsname{\let\PYG@it=\textit\def\PYG@tc##1{\textcolor[rgb]{0.25,0.50,0.56}{##1}}}
\expandafter\def\csname PYG@tok@s1\endcsname{\def\PYG@tc##1{\textcolor[rgb]{0.25,0.44,0.63}{##1}}}
\expandafter\def\csname PYG@tok@ge\endcsname{\let\PYG@it=\textit}
\expandafter\def\csname PYG@tok@gd\endcsname{\def\PYG@tc##1{\textcolor[rgb]{0.63,0.00,0.00}{##1}}}
\expandafter\def\csname PYG@tok@s\endcsname{\def\PYG@tc##1{\textcolor[rgb]{0.25,0.44,0.63}{##1}}}
\expandafter\def\csname PYG@tok@vg\endcsname{\def\PYG@tc##1{\textcolor[rgb]{0.73,0.38,0.84}{##1}}}
\expandafter\def\csname PYG@tok@ow\endcsname{\let\PYG@bf=\textbf\def\PYG@tc##1{\textcolor[rgb]{0.00,0.44,0.13}{##1}}}
\expandafter\def\csname PYG@tok@m\endcsname{\def\PYG@tc##1{\textcolor[rgb]{0.13,0.50,0.31}{##1}}}
\expandafter\def\csname PYG@tok@gh\endcsname{\let\PYG@bf=\textbf\def\PYG@tc##1{\textcolor[rgb]{0.00,0.00,0.50}{##1}}}
\expandafter\def\csname PYG@tok@c\endcsname{\let\PYG@it=\textit\def\PYG@tc##1{\textcolor[rgb]{0.25,0.50,0.56}{##1}}}
\expandafter\def\csname PYG@tok@nn\endcsname{\let\PYG@bf=\textbf\def\PYG@tc##1{\textcolor[rgb]{0.05,0.52,0.71}{##1}}}
\expandafter\def\csname PYG@tok@cs\endcsname{\def\PYG@tc##1{\textcolor[rgb]{0.25,0.50,0.56}{##1}}\def\PYG@bc##1{\setlength{\fboxsep}{0pt}\colorbox[rgb]{1.00,0.94,0.94}{\strut ##1}}}
\expandafter\def\csname PYG@tok@kn\endcsname{\let\PYG@bf=\textbf\def\PYG@tc##1{\textcolor[rgb]{0.00,0.44,0.13}{##1}}}
\expandafter\def\csname PYG@tok@mb\endcsname{\def\PYG@tc##1{\textcolor[rgb]{0.13,0.50,0.31}{##1}}}
\expandafter\def\csname PYG@tok@vi\endcsname{\def\PYG@tc##1{\textcolor[rgb]{0.73,0.38,0.84}{##1}}}
\expandafter\def\csname PYG@tok@si\endcsname{\let\PYG@it=\textit\def\PYG@tc##1{\textcolor[rgb]{0.44,0.63,0.82}{##1}}}
\expandafter\def\csname PYG@tok@nb\endcsname{\def\PYG@tc##1{\textcolor[rgb]{0.00,0.44,0.13}{##1}}}
\expandafter\def\csname PYG@tok@mo\endcsname{\def\PYG@tc##1{\textcolor[rgb]{0.13,0.50,0.31}{##1}}}
\expandafter\def\csname PYG@tok@cm\endcsname{\let\PYG@it=\textit\def\PYG@tc##1{\textcolor[rgb]{0.25,0.50,0.56}{##1}}}
\expandafter\def\csname PYG@tok@kt\endcsname{\def\PYG@tc##1{\textcolor[rgb]{0.56,0.13,0.00}{##1}}}
\expandafter\def\csname PYG@tok@sr\endcsname{\def\PYG@tc##1{\textcolor[rgb]{0.14,0.33,0.53}{##1}}}
\expandafter\def\csname PYG@tok@na\endcsname{\def\PYG@tc##1{\textcolor[rgb]{0.25,0.44,0.63}{##1}}}
\expandafter\def\csname PYG@tok@mf\endcsname{\def\PYG@tc##1{\textcolor[rgb]{0.13,0.50,0.31}{##1}}}
\expandafter\def\csname PYG@tok@vc\endcsname{\def\PYG@tc##1{\textcolor[rgb]{0.73,0.38,0.84}{##1}}}
\expandafter\def\csname PYG@tok@sh\endcsname{\def\PYG@tc##1{\textcolor[rgb]{0.25,0.44,0.63}{##1}}}
\expandafter\def\csname PYG@tok@ne\endcsname{\def\PYG@tc##1{\textcolor[rgb]{0.00,0.44,0.13}{##1}}}
\expandafter\def\csname PYG@tok@gt\endcsname{\def\PYG@tc##1{\textcolor[rgb]{0.00,0.27,0.87}{##1}}}
\expandafter\def\csname PYG@tok@nd\endcsname{\let\PYG@bf=\textbf\def\PYG@tc##1{\textcolor[rgb]{0.33,0.33,0.33}{##1}}}
\expandafter\def\csname PYG@tok@il\endcsname{\def\PYG@tc##1{\textcolor[rgb]{0.13,0.50,0.31}{##1}}}
\expandafter\def\csname PYG@tok@gp\endcsname{\let\PYG@bf=\textbf\def\PYG@tc##1{\textcolor[rgb]{0.78,0.36,0.04}{##1}}}
\expandafter\def\csname PYG@tok@err\endcsname{\def\PYG@bc##1{\setlength{\fboxsep}{0pt}\fcolorbox[rgb]{1.00,0.00,0.00}{1,1,1}{\strut ##1}}}
\expandafter\def\csname PYG@tok@se\endcsname{\let\PYG@bf=\textbf\def\PYG@tc##1{\textcolor[rgb]{0.25,0.44,0.63}{##1}}}
\expandafter\def\csname PYG@tok@sb\endcsname{\def\PYG@tc##1{\textcolor[rgb]{0.25,0.44,0.63}{##1}}}
\expandafter\def\csname PYG@tok@go\endcsname{\def\PYG@tc##1{\textcolor[rgb]{0.20,0.20,0.20}{##1}}}
\expandafter\def\csname PYG@tok@nt\endcsname{\let\PYG@bf=\textbf\def\PYG@tc##1{\textcolor[rgb]{0.02,0.16,0.45}{##1}}}
\expandafter\def\csname PYG@tok@s2\endcsname{\def\PYG@tc##1{\textcolor[rgb]{0.25,0.44,0.63}{##1}}}
\expandafter\def\csname PYG@tok@mh\endcsname{\def\PYG@tc##1{\textcolor[rgb]{0.13,0.50,0.31}{##1}}}
\expandafter\def\csname PYG@tok@kp\endcsname{\def\PYG@tc##1{\textcolor[rgb]{0.00,0.44,0.13}{##1}}}
\expandafter\def\csname PYG@tok@kr\endcsname{\let\PYG@bf=\textbf\def\PYG@tc##1{\textcolor[rgb]{0.00,0.44,0.13}{##1}}}
\expandafter\def\csname PYG@tok@cpf\endcsname{\let\PYG@it=\textit\def\PYG@tc##1{\textcolor[rgb]{0.25,0.50,0.56}{##1}}}
\expandafter\def\csname PYG@tok@sx\endcsname{\def\PYG@tc##1{\textcolor[rgb]{0.78,0.36,0.04}{##1}}}
\expandafter\def\csname PYG@tok@no\endcsname{\def\PYG@tc##1{\textcolor[rgb]{0.38,0.68,0.84}{##1}}}
\expandafter\def\csname PYG@tok@mi\endcsname{\def\PYG@tc##1{\textcolor[rgb]{0.13,0.50,0.31}{##1}}}
\expandafter\def\csname PYG@tok@o\endcsname{\def\PYG@tc##1{\textcolor[rgb]{0.40,0.40,0.40}{##1}}}
\expandafter\def\csname PYG@tok@ss\endcsname{\def\PYG@tc##1{\textcolor[rgb]{0.32,0.47,0.09}{##1}}}
\expandafter\def\csname PYG@tok@k\endcsname{\let\PYG@bf=\textbf\def\PYG@tc##1{\textcolor[rgb]{0.00,0.44,0.13}{##1}}}
\expandafter\def\csname PYG@tok@gs\endcsname{\let\PYG@bf=\textbf}
\expandafter\def\csname PYG@tok@cp\endcsname{\def\PYG@tc##1{\textcolor[rgb]{0.00,0.44,0.13}{##1}}}
\expandafter\def\csname PYG@tok@gr\endcsname{\def\PYG@tc##1{\textcolor[rgb]{1.00,0.00,0.00}{##1}}}
\expandafter\def\csname PYG@tok@sc\endcsname{\def\PYG@tc##1{\textcolor[rgb]{0.25,0.44,0.63}{##1}}}
\expandafter\def\csname PYG@tok@gi\endcsname{\def\PYG@tc##1{\textcolor[rgb]{0.00,0.63,0.00}{##1}}}
\expandafter\def\csname PYG@tok@ni\endcsname{\let\PYG@bf=\textbf\def\PYG@tc##1{\textcolor[rgb]{0.84,0.33,0.22}{##1}}}
\expandafter\def\csname PYG@tok@gu\endcsname{\let\PYG@bf=\textbf\def\PYG@tc##1{\textcolor[rgb]{0.50,0.00,0.50}{##1}}}
\expandafter\def\csname PYG@tok@nl\endcsname{\let\PYG@bf=\textbf\def\PYG@tc##1{\textcolor[rgb]{0.00,0.13,0.44}{##1}}}
\expandafter\def\csname PYG@tok@kc\endcsname{\let\PYG@bf=\textbf\def\PYG@tc##1{\textcolor[rgb]{0.00,0.44,0.13}{##1}}}

\def\PYGZob{\char`\{}
\def\PYGZcb{\char`\}}

\def\PYGZgt{\char`\>}

\makeatother

\begin{document}

\maketitle
\tableofcontents
\phantomsection\label{index::doc}

This document is a technical overview and discussion of our work, a protocol
for secure group messaging. By \emph{secure} we mean for the actual users i.e.
end-to-end security, as opposed to ``secure'' for irrelevant third parties.

Our work provides everything needed to run a messaging session between real
users on top of a real transport protocol. That is, we specify not just a key
exchange, but when and how to run these relative to transport-layer events; how
to achieve \emph{liveness properties} such as reliability and consistency, that are
time-sensitive and lie outside of the send-receive logic that cryptography-only
protocols often restrict themselves to; and offer suggestions for displaying
accurate (i.e. secure) but not overwhelming information in user interfaces.

We aim towards a general-purpose unified protocol. In other words, we'd prefer
to avoid creating a completely new protocol merely to support automation, or
asynchronity, or a different transport protocol. This would add complexity to
the overall ecosystem of communications protocols. It is simply unnecessary if
the original protocol is designed well, as we have tried to do.

That aim is not complete -- our full protocol system, as currently implemented,
is suitable only for use with certain instant messaging protocols. However, we
have tried to separate out conceptually-independent concerns, and solve these
individually using minimal assumptions even if other components make extra
assumptions. This means that many components of our full system can be reused
in future protocol extensions, and we know exactly which components must be
replaced in order to lift the existing constraints on our full system.

Our main implementation is a JavaScript library; an initial version is awaiting
integration into the MEGA web chat platform. We also have a Python reference
prototype that is semi-functional but omits real cryptographic algorithms. Both
our design and implementation need external review. To that end, this document
contains high-level descriptions of each, and we have also published source
code and API documentation under the free open source software AGPL-3 license.

In chapters 1 and 2, we explore topics in secure group messaging in general,
then apply these to our work and justify our design choices. In chapter 3, we
present our full protocol system, including our chosen separations of concerns
and how we achieve various security properties. In chapter 4, we describe our
implementation, including our software architecture, along with paradigms we
employ in lower-level utilities. In chapter 5, we detail our crytographic
algorithms and our packet format. Finally in chapter 6, we outline future work,
both for the immediate short term as well as long-term research objectives.
\newpage
\begin{DUlineblock}{0em}
\item[] © 2015 Mega Limited, Auckland, New Zealand.
\item[] \url{https://mega.nz/}
\end{DUlineblock}\vspace{-\baselineskip}

This work is licensed under the Creative Commons
Attribution-ShareAlike 4.0 International License. For the full text of the license in various formats, and other
details, see \url{https://creativecommons.org/licenses/by-sa/4.0/}

To give a non-legal human-readable summary of (and \emph{not} a substitute for) the
license, you are free to:
\begin{itemize}
\item {} 
Share -- copy and redistribute the material in any medium or format

\item {} 
Adapt -- remix, transform, and build upon the material

\end{itemize}

for any purpose, even commercially. The licensor cannot revoke these freedoms
as long as you follow the license terms:
\begin{itemize}
\item {} 
Attribution -- You must give appropriate credit, provide a link to the
license, and indicate if changes were made; but not in any way that suggests
the licensor endorses you or your use.

\item {} 
ShareAlike -- If you distribute your adaptions of the material, you must do
this under the same license as the original. You may not apply additional
restrictions when doing so, legal or technical.

\end{itemize}

No warranties are given. Other legal rights may extend (e.g. fair use and fair
dealing) or restrict (e.g. publicity, privacy, or moral rights) your permission
to use this material, outside of the freedoms given by this license.

\chapter{Background}
\label{background:background}\label{background::doc}\label{background:multi-party-encrypted-messaging-protocol}
This chapter is not about our work, but a general discussion of secure group
communications -- our model of what it is, and ``ideal'' properties that \emph{might}
be achieved. It is the longest chapter, so readers already familiar with such
topics may prefer to skip to the next one.

Often, lists of security properties can seem arbitrary, with technical names
that seem unrelated to each other. We take a more methodical approach, and try
to classify these properties within a general framework. To be clear, this is
neither formal nor precise, and our own project goals only focus on a subset of
these. Our motivation is to \emph{enumerate} all options from a \emph{protocol design}
perspective, for future reference and for comparison with other projects that
focus on a different subset. It offers some assurance that we haven't missed
anything, provides better understanding of relationships between properties,
and suggests natural separations for solving different concerns.

There is other work, previous and ongoing, that gives more precise treatments
of the topics below. We encourage interested readers to explore those for
themselves, as well as future research on classification and enumeration.

\section{Model and mechanics}
\label{background:model-and-mechanics}
First, we present an abstract conceptual model of a \emph{private group session} and
introduce some terminology. Secure communication systems generally consist of
the following steps:
\begin{enumerate}
\setcounter{enumi}{-1}
\item {} 
Identity (long-term) key validation.

\item {} 
Session membership change (e.g. establishment or optional termination).

\item {} 
Session communication.

\end{enumerate}

We only concern ourselves with (1) and (2). We assume that (0), a.k.a. the ``PKI
problem'', is handled by an external component that the application interacts
directly with, bypassing our components. Even if it is not handled safely, this
does not affect the \emph{functionality} of our components; but the application
should display a warning \footnote[1]{\sphinxAtStartFootnote%
For example, ``WARNING: the authenticity and privacy of this session
is dependant on the unknown validity of the binding \$key ↔
\$user'' or perhaps something less technical.
} and record this fact and/or have the user
complete that step retroactively.

A \textbf{session} (as viewed by a subject member, at a given time) is formed from a
set of transport \emph{packets}, which the protocol interprets logically as a set of
messages and a \emph{session membership}. Each \textbf{message} has a single \emph{author} and
a set of \emph{readers}; the \emph{message membership} is the union of these. Messages
have an order relative to each other, that may be represented (without loss of
information) as a set of \emph{parent messages} for each message; and relative to
the session boundary, i.e. when the subject decides to join and part.

Whilst part of a session, members may change the session membership, send new
messages to the current session membership, and receive these events from each
other. Events are only readable to those who are part of the session membership
of the event generator, when they generated it. For example, joining members
cannot read messages that were written before the author saw them join (unless
an old member leaks the contents to them, outside of the protocol). \footnote[2]{\sphinxAtStartFootnote%
We don't yet have a good model of what it should precisely mean to
\emph{rejoin} a session. This ``happens to work'' with what we've implemented, but
is not easily extensible to asynchronous messaging. Specifically, it's
unclear how best to consistently define the relative ordering of messages
across all members. We will explore this topic in more depth in the future.
}

On the transport level, we assume an efficient packet broadcast operation, that
costs time near-constant in the number of recipients, and bandwidth near-linear
(or less) in the number of recipients or the size of the packet. Note that the
\emph{sender} and \emph{recipients} of a transport packet are concepts distinct from the
author and readers of a logical message; our choice of terminology tries to
make this clear and unambiguous.

For completeness, we observe that a real system often unintentionally generates
``side channel'' information, beyond the purpose of the model. This includes the
time, space, energy cost of computation; the size, place of storage; the time,
size, route of communications; and probably more that we've missed. Often it is
impossible to avoid generating some of this information.

In summary, we've enumerated the broad categories of information in our model:
membership, ordering, content, and side channels. Next, we'll discuss and
classify the security properties we might want, then consider these properties
in the context of each of these categories.

\section{Security properties}
\label{background:security-properties}
In any information network, we produce and consume information. This could be
explicit (e.g. contents) or implicit (e.g. metadata). From this very general
observation, we can suggest a few fundamental security properties:
\begin{description}
\item[{\textbf{Efficiency}}] \leavevmode
Nobody should be able to cause anyone to spend resources (i.e. time, memory,
or bandwidth) much greater than what the initiator spent.

\item[{\textbf{Authenticity}}] \leavevmode
Information should be associated with a proof, either explicit or implicit,
that consumers may use to verify the truth of it.

\item[{\textbf{Confidentiality}}] \leavevmode
Only the intended consumers should be able to access and interpret the
information.

\end{description}

We'll consider authenticity and confidentiality as applied to the categories of
information we listed above. (Efficiency is more fiddly and we'll discuss it in
narrower terms later, applied to specific parts of our protocol system.)

Here is a summary of current known best techniques for achieving each property.
Though we don't try to achieve all of them in our protocol system, being aware
of them allows us to avoid decisions that destroy the possibility to achieve
them in the future.

\noindent\begin{tabulary}{\linewidth}{|L|L|L|}
\hline
\textsf{\relax } & \textsf{\relax 
Authenticity
} & \textsf{\relax 
Confidentiality
}\\
\hline
Existence
 & 
automatic
 & 
research topic
\\
\hline
Side channels
 & 
unneeded
 & 
research topic
\\
\hline
Membership
 & 
via crypto
 & 
research topic
\\
\hline
Ordering
 & 
via crypto
 & 
not explored yet
\\
\hline
Contents
 & 
via crypto
 & 
via crypto
\\
\hline\end{tabulary}

\begin{description}
\item[{Auth. of session existence}] \leavevmode
This is achieved automatically by authenticity of any of the other types; we
don't need to worry about it on its own.

\item[{Conf. of session existence}] \leavevmode
This is the hardest to achieve, and is an ongoing research topic. This is the
scenario where the user must hide the fact that they are merely \emph{using the
protocol}, even if the attacker knows nothing about any actual sessions. Not
only does it require confidentiality of all the other types of information,
but also obfuscation, steganography, and/or anti-forensics techniques.

\item[{Auth. of side channels}] \leavevmode
We don't care about the authenticity of something we didn't intend to
communicate in the first place.

\item[{Conf. of side channels}] \leavevmode
Still a research topic, this is a concern because it may be used to break
the confidentiality of other types of information.

\item[{Auth. of membership}] \leavevmode
There is some depth to this. The first choice is whether a distinct \emph{change
session membership} operation should exist outside of sending messages. ``Yes''
means that (e.g.) you can add someone to the session, and they will know this
(maybe a window will pop up on their side) even if you don't send them any
messages. ``No'' means that membership changes must always be associated with
an actual message that effects this change. This is up to the application;
though ``no'' is generally more suited for asynchronous messaging.

If ``yes'', we must consider \emph{entity aliveness} in our membership protocol.
This is the property that \emph{if} we complete the protocol successfully, \emph{then}
we are also sure that our peers have done the same thing. This requires a key
confirmation step from joining members, making the protocol last at least 2
rounds. If we don't consider this, then we may use shorter protocols, but
then our peers might not know that we changed the session membership until we
send them a message, which makes the ``yes'' choice less useful.

In both cases, authenticating (adding a proof of authorship to) messages is
not enough to verify membership, since packets may get dropped, perhaps even
against different recipients. We need to wait for all readers to send a reply
back to acknowledge their receipt. This is known as \emph{reliability} (when the
author checks it) or \emph{consistency} (when a reader checks the other readers).

\item[{Conf. of membership}] \leavevmode
This is more commonly called \emph{unlinkability} and is an ongoing research
topic. One major difficulty is that the information may be inferred from many
different sources, often implicit in the implementation or in the choice of
transport, and not explicit in the model. For example:
\begin{itemize}
\item {} 
It may be inferred from side channels, such as timing or packet size
correlations at the senders and recipients, or transport-defined headers
that contain routing information. So, we need routing anonymity and padding
or chaff mechanisms to protect against this line of attack.

\item {} 
It may be inferred from content, such as raw signatures or public keys.
So, we need confidential authentication mechanisms (defined later) as well
as obfuscated transport protocols to protect against this line of attack.

\end{itemize}

\item[{Auth. of ordering}] \leavevmode
The two types of ordering we identified earlier are: ordering of messages
within a session, and session boundary ordering relative to local events. The
latter is more commonly known as \emph{freshness}.

Other systems sometimes claim freshness based on some idea of an absolute
clock, but this requires trusting third-party infrastructure and/or the
user's local clock being correct. We prefer to avoid such approaches as it
makes the guarantees less clear.

Cryptographic guarantees are best; e.g., if we witness (in \emph{any order}) a
hash and its pre-image, then we are sure that the former was generated (i.e.
authored) \emph{after} the latter. To link remote events to our own time, we can
arrange for the pre-image to be derived from some unpredictable local event
such as generating a random nonce.

\item[{Conf. of ordering}] \leavevmode
This hasn't been explored explicitly in the wider literature, but could help
to break confidentiality of other types of information. It's out of scope for
us to consider it in more detail right now; sorry.

\item[{Auth. of contents}] \leavevmode
This is a straightforward application of cryptography. To be clear, this is
about proving that the author intended to send us the contents. Whether the
claims in it are true (including any implicit claim that the contents weren't
copied from elsewhere) is another matter. We'll refer back to this later.

\item[{Conf. of contents}] \leavevmode
This is a straightforward application of cryptography.

\end{description}

In conclusion, we've taken a top-down approach to identify security properties
that are direct high-level user concerns in a general \emph{private group session}.
We have not considered lower-level properties (e.g. \emph{contributiveness}, \emph{key
control}) here since they are only relevant to specific implementations; but we
will discuss such properties \emph{when and if} they might affect the ones above.

\section{Threat models}
\label{background:threat-models}
Attackers with different powers may try to break any of the above properties.
First, let's define and describe these powers:
\begin{description}
\item[{Active communications attack (on the entire transport)}] \leavevmode
This is the standard attack that all modern communications systems should
protect against -- i.e. a transport-level ``man in the middle'' who can inject,
drop, replay and reorder packets. Generally (and in practise mostly) channels
are bi-directional -- so we must assume that the attacker, if they are able
to target one member, then they are able to target all members, by attacking
the channel in the opposite direction.

Since this a baseline requirement, the powers defined below should be taken
to \emph{include} the ability to actively attack the transport during any attack.

\item[{Leak identity secrets (of some targets)}] \leavevmode
This refers to all of the secret material needed for a subject to establish a
new session, e.g. with someone that they've never communicated with before.
By definition, this is secret only to the subject.

\item[{Leak session secrets (of some targets)}] \leavevmode
This refers to all of the secret material needed for a subject to continue
participating in a session they're already part of. This may be secret to
only the subject, or shared across all members, or a combination of both.

This may \emph{include} identity secrets if they must be used to generate/process
messages or membership changes. Better protocols would \emph{not} need them, so
that they may be wiped from memory during a session, reducing the attack
surface. However, even in the latter case, session secrets still contain
\emph{entropy} from identity secrets; weaker threat models that exclude this are
better for modelling attacks on the RNG than on memory; see \phantomsection\label{background:id3}{\hyperref[references:ssek]{\crossref{{[}11SSEK{]}}}}.

\item[{Corrupt member (of some targets)}] \leavevmode
This may be either a genuine but malicious member, or a external attacker
that has exploited the running system of a genuine member. Unfortunately, it
is probably impossible to distinguish the two cases.

We include this for completeness as ``the worst case'', though it's unclear if
this is fundamentally different from many repeated applications of ``leak
session secrets''; more research is needed here. One difference could be that
under the corruption attack, there is no parallel honest instance \emph{and} the
attacker can observe secret computations that don't touch memory, e.g.
collecting, using, then immediately discarding entropy.

\end{description}

Now, we enumerate all the concrete things an attacker \emph{might} be able to do, by
applying these attacks to our model of session mechanics. For simplicity, we
only consider individual attacks; a full precise formal treatment will need to
consider multiple attacks across multiple sessions. In the following, the term
``target'' refers to the direct target of a secrets-leak or corruption, and the
term ``current'' refers to when that attack happens.
\begin{description}
\item[{Older sessions (i.e. already-closed)}] \leavevmode\vspace{-\baselineskip}\begin{itemize}
\item {} 
read session events (i.e. decrypt/verify messages and membership changes);
\footnote[3]{\sphinxAtStartFootnote%
It may be theoretically possible to restrict this only to messages
authored \emph{by non-targets} (excluding messages that the \emph{target} sends), and
likewise for membership changes. However, it's unlikely that the complexity
of any solution is worth the benefit, since (a) for every member, we'd need
to arrange that they can't derive the ability to decrypt their own messages
from their session secrets, and (b) even with this protection, the attacker
can still just compromise a second member to get the missing pieces.
} \footnote[4]{\sphinxAtStartFootnote%
The inability for an attacker to decrypt past messages is commonly
known as \emph{forward secrecy}. Currently are several slightly different formal
models for this, but the general idea is the same.
}

\item {} 
(participation is not applicable, since the session is already closed).

\end{itemize}

\item[{Current sessions (i.e. opened, not-yet-closed)}] \leavevmode\vspace{-\baselineskip}\begin{itemize}
\item {} 
read session events; \footnotemark[4]

\item {} 
participate as targets (i.e. auth/encrypt messages and initiate/confirm
membership changes; this includes making false claims or omissions in the
\emph{contents} of messages, such as receipt acknowledgements, which may break
other properties like authenticity of ordering or message membership, or
invariants on the application layer);

\item {} 
participate as non-targets (e.g. against the targets); \footnote[5]{\sphinxAtStartFootnote%
This attack is commonly known as \emph{key compromise impersonation}.
As with forward secrecy, there are slightly different models for this.
}

\item {} 
(these may apply differently to newer or older parts of the session).

\end{itemize}

\item[{Newer sessions (i.e. not-yet-opened)}] \leavevmode\vspace{-\baselineskip}\begin{itemize}
\item {} 
open/join sessions as targets, read their events and participate in them;

\item {} 
open/join sessions (etc.) as non-targets (e.g. invite a target or intercept
or respond to their invitations), read and participate in them; \footnotemark[5]

\item {} 
read session events, participate as targets, or as non-targets, in sessions
whose establishment was not compromised as in the previous points. \footnote[6]{\sphinxAtStartFootnote%
For example, the protocol may offer the ability for members to check
out-of-band, after establishment, that shared session secrets match up as
expected, and if so then be convinced that the session has full security
(i.e. session establishment was not tampered with) \emph{even if members know}
that identity secrets were already compromised before establishment.
}

\end{itemize}

\end{description}

Next, we'll discuss the unavoidable consequences of attacks using these powers,
and from this try to get an intuitive idea on the best thing we \emph{might} be able
to defend against:
\begin{description}
\item[{Active attack}] \leavevmode
Present cryptographic systems have security theorems that state that, when
implemented correctly and assuming the hardness of certain mathematical
problems, an attacker at this level cannot break the confidentiality and
authenticity of message contents, or the authenticity of membership -- and
therefore also anything that derives their security from these properties.

However, side channel attacks against the confidentiality of membership are
feasible; hiding this sufficiently well (and even defining models for all of
the side channels) is still a research problem, and it is not known what the
maximum possible protection is.

\item[{Leak identity secrets (and active attack)}] \leavevmode
By definition, the attacker may open/join sessions \emph{as targets}, then read
their events and participate in them. However, we may be able to prevent them
from doing anything else.

\item[{Leak session secrets (and active attack)}] \leavevmode
By definition, the attacker may participate \emph{as targets} in current sessions,
and read its events, for at least several messages into the future -- until
members mix in new entropy secret to the attacker. If session secrets also
include identity secrets, the attacker also gains the abilities mentioned in
the previous point. However, we may be able to prevent them from doing
anything else.

\item[{Corrupt insider (and active attack)}] \leavevmode
This is similar to the above, except that there is no chance of recovery by
members mixing in entropy, until after the corruption is healed.

\end{description}

As just discussed, under certain attacks we can't protect confidentiality, even
for actions of members not directly targeted by that attack. Such is the nature
of our group session model. But we \emph{could} try to protect a related property:
\begin{description}
\item[{\textbf{Confidential authenticity}}] \leavevmode
Only the intended consumers should be able to verify the information. (Of
course, attackers who break confidentiality, may choose to believe the
information even without being able to verify it.)

\end{description}

For every type of information where we want authenticity \emph{and} confidentiality,
we should also aim for confidential authenticity -- if the attacker should be
unable to read its contents, they should also be unable to verify \emph{anything}
about it. As noted earlier, this is useful not merely for its own sake, but is
essential if we want to protect the confidentiality of membership. Furthermore,
this property can be removed on a higher layer, e.g. an opt-in method to sign
all messages with public signature keys, but once lost it cannot be regained.
So it is safer to default to \emph{having} this property.

Against an active attacker, this means that verification must be executable
only by other members, i.e. depend on session secrets. Against a corrupt
insider, who is already allowed to perform verification, this means we must
choose a deniable or zero-knowledge authentication mechanism, so that they are
at least unable to pass this certainty-in-belief onto third parties.

\chapter{Topics}
\label{topics:topics}\label{topics::doc}
This chapter states the goals for our messaging protocol, the constraints we
chose, and discusses the issues that occur under these contexts.

\section{Security goals}
\label{topics:security-goals}
We do not try to provide confidentiality of session existence, membership, or
metadata, under any attack scenario. (For example, presently-known attacks that
break these include timing or other side channel analysis, or simply reading
the metadata generated by the normal operation of the transport.) However, we
try to avoid incompatibilities with any future systems that do provide those
properties, based on our knowledge of existing research and systems that
attempt or approximate these. \footnote[1]{\sphinxAtStartFootnote%
We do use an opportunistic exponential padding scheme that hides the
lowest bits of message length, but have not analysed this under a strong
adversarial model.
}

We aim to directly provide confidentiality of message content; authenticity of
membership, ordering, and content; and (later) a limited form of confidential
authenticity of ordering and content. An overview follows; we will expand on it
in more detail in subsequent chapters.

We achieve authenticity and confidentiality of message content, by using modern
standard cryptographic primitives with ephemeral session keys. The security of
these keys, and the authenticity of session membership and boundary ordering
(freshness), are achieved via an authenticated group key agreement protocol. We
do \emph{not} use timestamps.

The authenticity of message ordering is achieved through the authenticity of
message content, together with some rules that enforce logical consistency.
That is, someone who can authenticate message contents (either an attacker via
secrets leak or a corrupt insider) must still adhere to those rules.

The authenticity of message membership (including reliability, consistency) is
achieved through the authenticity of message ordering, together with some rules
that ensure liveness using timeout warnings and continual retries. This differs
from previous approaches \phantomsection\label{topics:id2}{\hyperref[references:mpom]{\crossref{{[}09MPOM{]}}}} \phantomsection\label{topics:id3}{\hyperref[references:gotr]{\crossref{{[}13GOTR{]}}}} which achieve those via a separate
cryptographic mechanism. We believe that our approach is a cleaner separation
of concerns, requiring \emph{no extra cryptography} beyond author authentication and
re-using basic concepts from time-tested reliability protocols such as TCP.

Our choice of mechanisms are intended to retain all these properties when under
an active attack on the communication transport. Under stronger attacks, we
have lower but still reasonable levels of protection:
\begin{description}
\item[{Under identity secrets leak against some targets (and active attack):}] \leavevmode\vspace{-\baselineskip}\begin{description}
\item[{Older sessions:}] \leavevmode\vspace{-\baselineskip}\begin{itemize}
\item {} 
Retain all relevant security properties.

\end{itemize}

\item[{Current sessions:}] \leavevmode\vspace{-\baselineskip}\begin{itemize}
\item {} 
Retain all relevant security properties until the next membership change;

\item {} 
{[}+{]} From the next change onwards, properties are as (a), since in our
system membership changes require identity secrets to execute.

\end{itemize}

\item[{Newer sessions (a):}] \leavevmode\vspace{-\baselineskip}\begin{itemize}
\item {} 
{[}x{]} Attacker can open/join sessions as the targets of the leak, and read
and participate those sessions;

\item {} 
Attacker \emph{cannot} open/join sessions as non-targets (those unaffected by
the leak), not even with the compromised targets;

\item {} 
Retain all relevant security properties, for sessions whose establishment
was not actively compromised.

\end{itemize}

\end{description}

\item[{Under session secrets leak against some targets (and active attack) (b):}] \leavevmode\vspace{-\baselineskip}\begin{description}
\item[{Older sessions:}] \leavevmode\vspace{-\baselineskip}\begin{itemize}
\item {} 
Retain all relevant security properties.

\end{itemize}

\item[{Current sessions:}] \leavevmode\vspace{-\baselineskip}\begin{itemize}
\item {} 
{[}+; partly x{]} Attacker can read session events;

\item {} 
{[}+; partly x{]} Attacker can participate as targets;

\item {} 
Attacker \emph{cannot} participate as non-targets.

\end{itemize}

\item[{Newer sessions:}] \leavevmode\vspace{-\baselineskip}\begin{itemize}
\item {} 
{[}+{]} as (a); our session secrets unfortunately include identity secrets.

\end{itemize}

\end{description}

\item[{Under insider corruption (and under active attack):}] \leavevmode
As (b) except that entries marked imperfect cannot be improved upon. More
specifically, the properties we \emph{do} retain are:
\begin{itemize}
\item {} 
Some limited protection against false claims/omissions about ordering;

\item {} 
(future work) Retain confidential authenticity of ordering and content.

\end{itemize}

Note that these apply to all lesser attacks too; we mention them explicitly
here so that this section is less depressing.

\end{description}

\vspace{-\baselineskip}\begin{DUlineblock}{0em}
\item[] {[}x{]} unavoidable, as explained in the previous chapter.
\item[] {[}+{]} imperfect, theoretically improvable, but we have no immediate plans to.
\end{DUlineblock}\vspace{-\baselineskip}

\section{Distributed systems}
\label{topics:distributed-systems}\label{topics:id4}
Security properties are meant to detect \emph{incorrect behaviour}; but conflicts in
naively-implemented distributed systems can happen \emph{even when} everyone behaves
correctly. If we don't explicitly identify and resolve these situations as \emph{not
a security problem}, but instead allow our security mechanisms to warn or fail
in their presence, we reduce the usefulness of the latter. In other words, a
warning which is 95\% likely to be a false positive, is useless information and
a very bad user experience that may push them towards less secure applications.

Generally in a distributed system, events may happen concurrently, so ideally a
causal order (directed acyclic graph) rather than a total order (line) should
be used to represent the ordering of events. We do this for messages, and this
component of our system may be directly reused for asynchronous messaging.

However, group key agreements (GKAs), which is our membership change mechanism,
historically have not been developed with this consideration in mind. A direct
solution would define the result of merging two concurrent operations, and give
an algorithm for both sides of the fork to execute, to reach this result state.
But we could not find literature that even mentions this problem, and we are
unsure how to begin to approach it. Based on our moderate experience, it seems
feasible at least that any solution would be highly specific to the GKA used,
limiting our future options for replacing cryptographic components.

For now, we give up causal ordering for membership operations, and instead
enforce a linear ordering for them. To make this easier, we restrict ourselves
to a group transport channel, that takes responsibility for the delivery of
channel events (messages and membership changes), reliably and in a consistent
order. \footnote[2]{\sphinxAtStartFootnote%
For example, XMPP MUC would be suitable for this purpose, since one
single server keeps a consistent order for the channel. In IRC, there may
be multiple servers that opportunistically forward messages from clients
to each other, without trying to agree on a consistent order.
} We do not \emph{assume} this; we detect any deviation from it and
notify the user, but our system is efficient if the transport is honest.

Beyond this, there are several more non-security distributed systems issues,
that relate to the integration of cryptographic logic in the context of a group
transport channel. These situations all have the potential to corrupt the
internal state consistency of a naive implementation:
\begin{itemize}
\item {} 
Two members start different operations concurrently (as mentioned above);

\item {} 
Two members try to complete an operation concurrently, in different ways;
e.g. ``send the final packet'' vs. ``send a request to abort'';

\item {} 
A user enters the channel during an operation;

\item {} 
A user leaves the channel during an operation;

\item {} 
A member starts an operation and sends the first packet to the other channel
members \(M\), but when others receive it the membership has changed to
\(M'\), or there was another operation that jumped in before it;

\item {} 
Different operation packets (initial or final) are decodeable by different
members, some of which are not part of the cryptographic session. If we're
not careful, they will think different things about the state of the session,
or of others' sessions;

\item {} 
Any of the above things could happen at the same time.

\end{itemize}

We must design graceful, low-failure-rate solutions for all of them. Individual
solutions to each of these are fairly straightforward, but making sure that
these interact with each other in a sane way is more complex. Then, there is
the task of describing the intended behaviour \emph{precisely}. Only when we have a
precise idea on what is \emph{supposed} to happen, can we construct a concrete
system that isn't fragile, i.e. require mountains of patches for corner cases
ignored during the initial hasty naive implementations.

\section{User experience}
\label{topics:user-experience}
Independently of any actual attack or security warning, the distributed nature
of our system requires us to consider how to represent \emph{correct} information to
users. Displaying inaccurate or vague information is a security risk \emph{even
without an attacker} because it can lead the user to believe incorrect things.

Here, we give an overview of these issues and our suggested solutions for them.
Avoiding any of these topics is always an option, which case the application
will look like -- \emph{and be as insecure as} -- existing applications that do the
same.
\begin{description}
\item[{Real parents of a message}] \leavevmode
Some messages may not be displayed immediately below the one(s) that they are
actually sent after, i.e. that the author saw when sending it.

Our suggestion: (a) allow the user to select a message (e.g. via mouse click,
long press or keyboard) upon which all non-ancestors are grayed out; and (b)
annotate the messages whose parents are not equal to the set \{the preceding
message in the UI\}, as a hint for the user to perform the selection.

\item[{Messages sent before a membership change completes, but received afterwards}] \leavevmode
Obviously, this message has a different membership from the current session,
and it would be wrong not to display this difference.

Our suggestion: (a) when an operation completes, issue a UI notice about it
inline in the messages view; (b) allow the user to select a message to see
its membership, instead of trying to infer it from the session membership and
any member change notices; and (c) annotate such messages as a hint for the
user to perform the selection.

\item[{Progress and result of a membership change operation}] \leavevmode
If the user starts an operation then immediately sends a message, this is
still encrypted to the \emph{old} membership. Unless we explicitly make it clear
that operations take a finite time, they may not realise this.

Our suggestion: issue UI notices inline in the messages view, when the user
proposes an operation and when it is rejected, is accepted (starts), fails or
succeeds; or (optionally) also when \emph{others'} operations are rejected, are
accepted, fail or succeed.

\item[{Messages received out-of-order}] \leavevmode
Some messages are sent, but the sent-later ones are received earlier.

Our suggestion: simply ignore the messages that are received too early, until
the missing gaps are filled. This might seem counter-intuitive, but there are
many reasons that this is the best behaviour, discussed in \phantomsection\label{topics:id6}{\hyperref[references:msg\string-2oo]{\crossref{{[}msg-2oo{]}}}}. There
are some other options, but we believe these are all strictly worse.

\item[{Messages not yet acknowledged by all of its intended readers}] \leavevmode
Here, we are unsure if everyone received what we sent, or received the same
messages that we received from others.

Our suggestion: (a) allow the user to select a message to see who has not yet
acknowledged it, out of its membership; (b) annotate such messages as a hint
for the user to perform the selection, after a grace timeout because it's
impossible to satisfy this immediately; and optionally (c) show a progress
meter for this condition for every message we send.

\item[{Users not responding to heartbeats}] \leavevmode
This helps to detect transports dropping our messages.

Our suggestion: in the users view, gray out expired users.

\end{description}

A more detailed discussion of these topics is given at \phantomsection\label{topics:id7}{\hyperref[references:msg\string-hci]{\crossref{{[}msg-hci{]}}}}.

\chapter{Protocol}
\label{protocol::doc}\label{protocol:protocol}
Here we present an overview of our protocol, suggested runtime data structures
and algorithms, with references to other external documents for more specific
information about each subtopic.

\section{Session overview}
\label{protocol:session-overview}
A session is a local process from the point of view of \emph{one member}. We don't
attempt to reason about nor represent an overall ``group'' view of a session; in
general such views are not useful for \emph{accurately} modelling, or implementing,
security or distributed systems.

A session, time-wise, contains a linear sequence of \emph{session membership change}
operations, that (if and when completed successfully) each create subsessions
of static membership, where members may send messages to each other.

Linearity is enforced through an acceptance mechanism, which is applied to the
packets where members try to start or finish an operation. The sequence is also
context-preserving, i.e. it is guaranteed that no other operations complete
\emph{between} when a member proposes one, and it being accepted by the group.

Each accepted operation \(G\) is initialised from (a) previous state (the
result of the previous operation, or a null state if the first) that encodes
the current membership \(M'\) and any cryptographic values that \(G\)
needs such as ephemeral keys; and (b) the first packet of the operation, sent
by a member of \(M'\), that defines the intended next membership \(M\).
While \(G\) is ongoing, members may send messages in the current subsession
as normal, i.e. to \(M'\). \footnote[1]{\sphinxAtStartFootnote%
Our first group key agreement implementation did not enforce atomic
operations. This caused major problems when users would leave the channel
at different times, e.g. if they disconnect or restart the application,
since their GKA components would see different packets and reach different
states. With atomic operations and our transport integration rules, an
inconsistent state is only reached if the transport (e.g. chat server)
behaves incorrectly. One of our goals is, that security warnings fire
\emph{only} when there is \emph{actually} (or \emph{likely} to be) a problem.
}

\(G\) may finish with success, upon which we atomically change to a new
subsession for \(M\), and store the result state for the next operation; or
with failure, upon which we destroy all temporary state related to \(G\),
and continue using the existing subsession with membership \(M'\).

After a successful change, the now-previous subsession (with membership
\(M'\)) enters a shutdown phase. This happens concurrently and
independently of other parts of the session, such as messaging in the new
subsession or subsequent membership change operations on top of \(M\).

\section{Group key agreement}
\label{protocol:group-key-agreement}
Our group key agreement is composed of the following two subprotocols:
\begin{itemize}
\item {} 
a group DH exchange to generate a shared ephemeral secret encryption key, for
confidentiality;

\item {} 
a custom group key distribution protocol for per-member ephemeral public
signature keys, for authenticity (of session membership) and freshness,
piggy-backed onto the same packets as the group DH packets.

\end{itemize}

The keys are used to read/write messages in the subsession created by the
operation. Identity secrets are \emph{not} needed for this, but they are needed for
participating in further membership changes (i.e. creating new subsessions).

For the overall protocol, the number of communication rounds is \(O(n)\) in
the number of members. The average size of GKA packets is also \(O(n)\).
More modern protocols have \(O(1)\) rounds but retain \(O(n)\) packet
size. However, our protocol is a simple first approach using elementary
cryptography only, which should be easier to understand and review.

This component may be upgraded or replaced independently of the other parts of
our protocol system. For example, more modern techniques make the key agreement
itself deniable via a zero-knowledge proof. We have skipped that for now to
solve higher-priority concerns first, and because implementing such protocols
correctly is more tricky. This is discussed further in {\hyperref[future\string-work::doc]{\crossref{\DUrole{doc}{Future work}}}}.

\section{Transport integration}
\label{protocol:transport-integration}\label{protocol:id2}
We have the initial packet of each operation, reference the final packet of the
previous finished operation (or null if the first). Likewise, the final packet
of each operation references (perhaps implicitly, if this is secure) a specific
initial packet. The concurrency resolver simply accepts the earliest packet in
the channel with a given parent reference, and rejects other such packets. We
also define a \emph{chain hash} built from the sequence of accepted packets, that
members verify the consistency of after each operation finishes.

The advantage of this \emph{implicit} agreement is that it has zero bandwidth cost,
generalises to \emph{any} membership change protocol, and does not depend on the
liveness of a particular member as it would if we had an explicit leader.

There are further details and cases in both the algorithm and implementation.
For example, we have additional logic to handle new members who don't know the
previous operation, and define a general method to authenticate the parent
references. We also arrange the code carefully to behave correctly for 1-packet
operations, where the packet acts both as an initial and a final packet in our
definitions above.

To cover the other {\hyperref[topics:distributed\string-systems]{\crossref{\DUrole{std,std-ref}{Distributed systems}}}} cases, we also have a system of
rules on how to react to various channel and session events. These work roughly
along these principles:
\begin{itemize}
\item {} 
Every operation's target members (i.e. joining members and remaining members)
must all be in the channel to see the first packet (otherwise it is ignored)
and remain there for the duration of the operation (otherwise it auto-fails).

\item {} 
Members that leave the channel are automatically excluded from the session,
and vice versa. There are subrules to handle events that conflict with this
auto-behaviour, that might occur before those behaviours are applied.

\item {} 
We never initiate membership operations to exclude ourselves. When we want to
part the session, we initiate a shutdown process on the subsession, wait for
it to finish, then leave the channel. When others exclude us, we wait for
them to kick us from the channel, if and after the operation succeeds. Either
way, we switch to a ``null/solo'' subsession only \emph{after} leaving the channel.

\end{itemize}

For full details of our agreement mechanism, our transport integration rules,
and the rationale for them, along with more precise descriptions of our model
for a general \(G\) and of a group transport channel, see \phantomsection\label{protocol:id3}{\hyperref[references:msa\string-5h0]{\crossref{{[}msa-5h0{]}}}}.

\section{Message ordering}
\label{protocol:message-ordering}
As discussed {\hyperref[topics:distributed\string-systems]{\crossref{\DUrole{std,std-ref}{previously}}}} and elsewhere \phantomsection\label{protocol:id4}{\hyperref[references:msg\string-2i0]{\crossref{{[}msg-2i0{]}}}}
messages may be sent concurrently even in a group transport channel, and so we
represent the transcript of messages as a cryptographic causal order. We take
much inspiration from \href{https://git-scm.com/}{Git} and OldBlue \phantomsection\label{protocol:id5}{\hyperref[references:oblu]{\crossref{{[}12OBLU{]}}}} for our ideas.

Every message has an explicit set of references to the latest other messages
(``parents'') seen by the author when they wrote the message. These references
are hashes of each packet, which require no extra infrastructure to generate or
resolve. When we decrypt and verify a packet, we verify the author of these
references as well. This allows us to ignore the order of packet receipt, and
instead construct our ordering by following these references. If we receive a
packet out-of-order, i.e. if we haven't yet received all of its parents, we
simply defer processing of it until we have received them.

References must at least be second-preimage-resistant, with the pre-image being
some function of the \emph{full} verified-decrypted referent message (i.e. content,
parents, author and readers), so that all members interpret them consistently.

Our definition based on hashing packet ciphertext, together with using a shared
group encryption key, guarantees the above property for our case. However to be
precise, it is important to note that such references are only claims. Their
truth is susceptible to lying; the claimant may:
\begin{itemize}
\item {} 
make false claims, i.e. refer to messages they haven't seen; second pre-image
resistance gives \emph{some} protection here, but an attacker could e.g. reuse a
hash value that they saw from another member;

\item {} 
make false omissions, i.e. not refer to messages that they have seen.

\end{itemize}

We have rules that enforce some logical consistency across references:
\begin{itemize}
\item {} 
a message's parents must form an anti-chain, i.e. none of these parents may
directly or indirectly (via intermediate messages) reference each other;

\item {} 
an author's own messages must form a total order (line).

\end{itemize}

This gives some protection against arbitrary lies, but it is still possible to
lie within these constraints. Nevertheless, we omit protection for the latter,
since we believe that there is no benefit for an attacker to make such lies,
and that the cost of any solution would not be worth the minor extra security.

For a more detailed exploration, including tradeoffs of the \emph{defer processing}
approach to strong ordering, and ways to calculate references to have better
resistance against false claims, see \phantomsection\label{protocol:id6}{\hyperref[references:msg\string-2o0]{\crossref{{[}msg-2o0{]}}}}.

\section{Reliability and consistency}
\label{protocol:git}\label{protocol:reliability-and-consistency}
Due to our strong ordering property, we can interpret parent references as an
implicit acknowledgement (``ack'') that the author received every parent. Based
on this, we can ensure end-to-end reliability and consistency. We take much
inspiration from the core ideas of \href{https://en.wikipedia.org/wiki/Transmission\_Control\_Protocol}{TCP}.

We require every message (those we write, \emph{and} those we read) to be acked by
all (other) readers; if we don't observe these within a timeout, we warn the
user. We may occasionally resend the packets of those messages (the subjects of
such warnings), including those authored by others. Resends are all based on
implicit conditions; we have no explicit resend requests as in OldBlue.

To ensure we ack everything that everyone wrote, we also occassionally send
acks automatically outside of the user's control. Due to strong ordering, acks
are transitive (i.e. implicitly ack all of its ancestors) and thus auto-acks
can be delayed to ``batch'' ack several messages at once and reduce volume.

We develop some extra details to avoid perpetual reacking-of-acks, yet ensure
that the final messages of a session, or of a busy period within a session, are
actually fully-acked. We also include a formal session shutdown process.

For a more detailed exploration, including resend algorithms, timing concepts,
different ack semantics, why we must have end-to-end authenticated reliability
instead of ``just using TCP'', the distinction between consistency and consensus,
and more, see \phantomsection\label{protocol:id7}{\hyperref[references:msg\string-2c0]{\crossref{{[}msg-2c0{]}}}}.

\section{Message encryption}
\label{protocol:message-encryption}\label{protocol:tcp}
For now, message encryption is very simple. Each subsession has a constant set
of keys (the output of the group key exchange) that are used to authenticate
and encrypt all messages in it -- one encryption key shared across all members,
and one signature key for each member, with the public part shared with others.

Every message is encrypted using the shared encryption key, then signed by the
author using their own private signature key. To decrypt, the recipient first
verifies the signature, then decrypts the ciphertext.

These are constant throughout the session, so that if the shared encryption key
is broken, the confidentiality of message content is lost. In the future, we
will experiment with implementing this component as a forward secrecy ratchet.
Note that we already have forward secrecy \emph{between} subsessions.

There is also the option to add a weak form of deniability, where authenticity
of message contents are deniable, but authenticity of session participation is
not. This is essentially the group analogue of how deniability is achieved in
OTR \phantomsection\label{protocol:id8}{\hyperref[references:otr\string-spec]{\crossref{{[}OTR-spec{]}}}}, and has equivalent security. (As mentioned before, making the
group key agreement itself deniable is stronger, but more complex to achieve.)
These directions are discussed further in {\hyperref[future\string-work::doc]{\crossref{\DUrole{doc}{Future work}}}}.

\chapter{Engineering}
\label{engineering::doc}\label{engineering:engineering}
In this chapter we describe the reference implementation of our protocol
system, its architecture and core engineering concepts.

For a distributed system, the only \emph{requirement} for it to work is that the
protocol is well-defined. However, protocols that support advanced features or
give strong guarantees are inherently more complex. Some people may argue that
the complexity is not worth the benefit, especially if it includes new ideas
untested by existing protocols. Our hope is that a high-level description of an
\emph{implementation} will help moderate this over-estimation of the complexity.
Another hope is that it will make audits and code reviews easier and faster.

\section{Public interface}
\label{engineering:public-interface}
We begin by observing that any implementation of any communications system must
(a) do IO with the network; and (b) do IO with the user. In our system, we
define the following interfaces for these purposes:
\begin{description}
\item[{\code{Session} (interface)}] \leavevmode
This models a group messaging session as described in {\hyperref[background::doc]{\crossref{\DUrole{doc}{Background}}}}. It
defines an interface for higher layers (e.g. the UI) to interact with, which
consists of (a) one input method to send a message or change the session
membership; (b) one output mechanism for the user to receive session events
or security warnings; and (c) various query methods, such as to get its
current membership, the stream of messages accepted so far, etc. \footnote[1]{\sphinxAtStartFootnote%
We do not define a lower (transport) interface in \code{Session} since
implementations or subtypes may require a \emph{particular} transport; we leave
it to them to define what that is. For example, \code{HybridSession} requires
a \code{GroupChannel} which makes it unsuitable for asynchronous messaging;
but another subtype of \code{Session} might support that.
}

\item[{\code{GroupChannel} (interface) {[}\${]}}] \leavevmode
This represents a group transport channel. It defines an interface for higher
layers (e.g. a \code{Session}) to interact with, which consists of (a) one input
method to send a packet or change the channel membership; (b) one output
mechanism to receive channel events; and (c) various query methods. Actions
may be ignored by the transport or satisfied exactly once, possibly after we
receive further channel events not by us. The transport is supposed to send
events to all members in the same order, but members must verify this.

\end{description}

\vspace{-\baselineskip}\begin{DUlineblock}{0em}
\item[] {[}\${]} This component is specific to instant or synchronous messaging; ones
\emph{not} marked with this may be reused in an asynchronous messaging system.
\end{DUlineblock}\vspace{-\baselineskip}

\code{Session} represents a logical view from one user; there is no distinction
between ``not in the session'' vs. ``in the session, and we are the only member''.
By contrast, \code{GroupChannel} is our view of an external entity (the channel),
and the analogous concepts for channel membership \emph{are} distinct.

So to use our library, an external application must:
\begin{enumerate}
\item {} 
Implement \code{GroupChannel} for the chosen transport, e.g. XMPP.

\item {} 
Construct an instance of \code{HybridSession} (see below), passing an instance
of (1) to it along with any other configuration options it wants.

\item {} 
Hook the application UI into the API provided by \code{Session}. The transport
layer may be ignored completely, since that is handled by our system.

\end{enumerate}

The last step may involve a lot of work if the application UI is too tightly
coupled with the specifics of a particular protocol. But there is no way around
this; a secure messaging protocol deals with concepts that are fundamentally
different from insecure transport protocols, and we see this already by the
difference between session and channel membership. Hopefully whoever does this
work will architect their future software with greater foresight.

The remainder of this document details the internals of our implementation; but
knowledge beyond this point is not necessary merely to \emph{use} our system.

\section{Session architecture}
\label{engineering:session-architecture}
\code{HybridSession} {[}\${]} is our main (and currently only) \code{Session}
implementation. It contains several internal components:
\begin{itemize}
\item {} 
a \code{GroupChannel}, a transport client for communication with the network;

\item {} 
a concurrency resolver, to gracefully prevent membership change conflicts;

\item {} 
a component {[}*{]} that manages and runs membership operations and proposals;

\item {} 
two components for the current and previous subsession; each contains:
\begin{itemize}
\item {} 
a message encryptor/decryptor {[}*{]} for communicating with the session;

\item {} 
a transcript data structure to hold accepted messages in the correct order;

\item {} 
various liveness components to ensure reliability and consistency.

\end{itemize}

\end{itemize}

\code{HybridSession} itself handles the various transport integration cases;
creates and destroys subprocesses to run membership operations; and manages
membership changes that are initiated by the local user, that require more
tracking such as retries in the case of transport hiccups, etc.

The receive handler roughly runs as follows. For each incoming channel event:
\begin{enumerate}
\item {} 
if it is a channel membership change, then react to as part of
{\hyperref[protocol:transport\string-integration]{\crossref{\DUrole{std,std-ref}{Transport integration}}}};

\item {} 
else, if it is a membership operation packet:
\begin{itemize}
\item {} 
if it is relevant to the concurrency resolver, pass it to that, which may
cause an operation to start or finish (with success or failure);

\item {} 
if an operation is ongoing, pass it to the subprocess running that;

\item {} 
else reject the packet (i.e. don't queue it for another try later).

\end{itemize}

\item {} 
else, try to verify-decrypt it as a message in the current subsession;
\begin{itemize}
\item {} 
if the packet verifies but fails to be accepted into the transcript due
to missing parents, put it in a queue \emph{specific to this subsession}, to
try this process again later (and similarly for the next case);

\end{itemize}

\item {} 
else, try to verify-decrypt it as a message in the previous subsession;

\item {} 
else, put it on a queue, to try this process again later, in case it was
received out-of-order and depends on missing packets to decrypt.

\end{enumerate}

The components that deal directly with cryptography are marked {[}*{]} above. These
may be improved independently from the others, and from \code{HybridSession}. We
may also replace the cryptographic primitives within each component -- e.g. DH
key exchange, signature schemes, hash functions and symmetric ciphers -- as
necessary, based on the recommendations of the cryptography community.

For more technical details, see our API documentation \phantomsection\label{engineering:id2}{\hyperref[references:mpenc\string-api]{\crossref{{[}mpenc-api{]}}}}.

\section{Internal components}
\label{engineering:internal-components}\begin{description}
\item[{\code{ServerOrder} {[}\${]}}] \leavevmode
The concurrency resolver, used by \code{HybridSession} to enforce a consistent
and context-preserving total ordering of membership operations. It tracks the
results of older operations, whether we're currently in an operation, and
decides how to accept/reject proposals for newer operations.

\item[{\code{Greeter}, \code{Greeting} (interface) {[}\${]}}] \leavevmode
\code{Greeting} represents a multi-packet operation. It defines an interface
with a packet-based transport consisting of (a) one input method to receive
data packets; (b) one output mechanism to send data packets; and (c) various
query methods, such as to get a \code{Future} for the operation's result, a
reference for the previous operation if there was one, the intended next
membership, etc. Typically, this may be implemented as a state machine.

\code{Greeter} is a factory component for new \code{Greeting} instances, defined as
an interface used by \code{HybridSession} that consists of some limited codec
methods for initial/final packets of a group key agreement. Implementations
of these methods may reasonably depend on state, such as the result of any
previous operation, data about operations proposed by the local user but not
yet accepted by the group, or a reference to an ongoing \code{Greeting}.

\item[{\code{SessionBase}}] \leavevmode
This is a partial \code{Session} implementation, for full implementations to
build around (as \code{HybridSession} does). It enforces properties such as
strong message ordering, reliability, and consistency, based on information
from message parent references and using some of the components below.

The component provides an interface with a packet-based transport consisting
of (a) one input method to receive data packets; (b) one output mechanism to
send data packets; and an interface with the UI consisting of (c) one output
mechanism for the user to receive notices; (d) various action methods for the
user to use, such as sending messages and ending the session; and (e) various
query methods similar to those found in \code{Session}.

Unlike \code{Session} (a), we make no attempt to simplify \code{SessionBase} (d) to
make it ``nice to use''. The functionality is quite low-level and may change in
the future; it is not meant for external clients of our system.

\end{description}

Everything from here on are components of \code{SessionBase}; \code{HybridSession}
does not directly interact with them (except \code{MessageLog}).
\begin{description}
\item[{\code{MessageSecurity} (interface)}] \leavevmode
This defines an interface for the authentication and encryption of messages.
The interface is flexible enough to allow implementations to generate new
keys based on older keys, and to implement automatic deletion rules for some
of those keys as they age further.

\item[{\code{Transcript}, \code{MessageLog}}] \leavevmode
These are append-only data structures that hold messages in causal order.

\code{Transcript} holds a causal ordering of all messages, including non-content
messages used for flow control and other lower-level concerns. It provides
basic query methods, and graph traversal and recursive merge algorithms. (The
latter is for aiding future research topics, and directly used yet. It may be
omitted in a time-constrained pragmatic reimplementation of our system.)

\code{MessageLog} is a \emph{user-level} abstraction of \code{Transcript}; it linearises
the underlying causal order for UX purposes, aggregates multiple transcripts
together (from multiple subsessions), and filters out non-content messages
whilst retaining causal ordering.

\item[{\code{FlowControl}}] \leavevmode
This defines an interface that \code{SessionBase} consults on liveness issues,
such as when to resend messages, how to handle duplicate messages, how to
react to packets that have been buffered for too long, etc. The interface is
designed to support using the same component across several \code{SessionBase}
instances, in case one wishes to make decisions based on all of their states.
The interface is private for the time being, since it is a bit unstructured
and may be changed later to fix this and other imperfections.

\item[{\code{ConsistencyMonitor}}] \leavevmode
This tracks expected acknowledgements for abstract items, and issues warnings
and/or tries to recover, if they are not received in a timely manner. It is
used by \code{SessionBase} and (in the future) \code{ServerOrder}.

\item[{\code{PresenceTracker}}] \leavevmode
This tracks our and others' latest activity in a session, and issues warnings
if these expire. This helps to detect drops by an unreliable transport or
malicious attacker. It can send out heartbeats to prevent or recover from
such situations, but this is optional since it has some bandwidth cost.

\end{description}

\section{Utilities}
\label{engineering:utilities}
Our protocol system is built from components that act as independent processes,
that react to inputs and generate outputs similar to the actor model. We build
up a relatively simple framework for this intra-process IO, based on some
low-level utilities. We'll talk about these first.

\subsection{Low-level}
\label{engineering:low-level}
For an input mechanism into a component that is decoupled from the source, we
simply use a function, since this exists in all major languages, and already
has the property that the callee doesn't know who the caller is.

For an output mechanism from a component that is decoupled from the target, we
use a synchronous publish-subscribe pattern. There are other options; the main
reason we choose this is that \emph{how} we consume inputs (of a given type) changes
often. For example: each new message adds a requirement that we do some extra
things on future messages; in trial decryption, the set of possible options
changes; etc. Pub-sub is ideal for these cases: we can subscribe new consumers
when we need to, and define their behaviour and cancellation conditions close
together in the source code.

By contrast, other intra-process IO paradigms (e.g. channels) are mostly built
around single consumers. Here, we'd have to collect all possible responses into
the consumer, then add explicit state to control the activation of specific
responses. This causes related concerns to be separated too much, and unrelated
concerns to be grouped together too much, and the mechanisms for doing this are
less standardised across common libraries.

By \emph{synchronous} we mean that the publisher executes subscriber callbacks in
its own thread. There are reentrancy issues around this \footnote[2]{\sphinxAtStartFootnote%
\emph{Reentrant publish} is when callbacks cause the producer to produce
new items \emph{whilst} they are being run. This can cause unintended behaviour,
sometimes called an \emph{interleaving hazard}, and is usually considered a bug.
See also \emph{\S{}13.1. Sequential Interleaving Hazards} in \phantomsection\label{engineering:id5}{\hyperref[references:robo]{\crossref{{[}06ROBO{]}}}}.

\emph{Reentrant subscribe/cancel} is when subscriptions for the current producer
are modified \emph{whilst} we are running the callbacks for one of its items.
The behaviour here must be precisely defined by the pub-sub system. For
example, web DOM events define that \href{https://developer.mozilla.org/en-US/docs/Web/API/EventTarget/removeEventListener\#Notes}{cancels take affect from the current
run}, but \href{https://developer.mozilla.org/en-US/docs/Web/API/EventTarget/addEventListener\#Adding\_a\_listener\_during\_event\_dispatch}{subscribes only take effect from the next run}.
}, but in our
simple usage it makes reasoning about execution order more predictable, and
means that we have no dependency on any specific external execution framework.

For long-running user operations, we use \code{Future}s, which is the standard
utility for this sort of asynchronous ``function call''-like operation, that is
expected to return some sort of response. In our system, a common pattern is
for a \code{Future}`s lifetime to include several IO rounds between components.

We chose to implement our own utilities for some of these things, to define
them in a more abstract style that is inspired from functional programming
languages. This allows us to write higher-order combinators, so that we can
express complex behaviours more concisely and generally.
\begin{description}
\item[{\code{Observable}}] \leavevmode
A pair of functions (publish, subscribe) and some mutable tracking state,
used to produce and consume items. The producer creates an instance of this,
keeps (publish) private and gives (subscribe) to potential consumers. In a
language that supports polymorphic types, we would have the following type
definitions, written in Scala-like pseudocode:

\begin{Verbatim}[commandchars=\\\{\}]
\PYG{k}{type} \PYG{k+kt}{Cancel}             \PYG{o}{=} \PYG{o}{(}\PYG{o}{)} \PYG{k}{=\PYGZgt{}} \PYG{n+nc}{Boolean}
\PYG{k}{type} \PYG{k+kt}{Subscribe}\PYG{o}{[}\PYG{k+kt}{T}, \PYG{k+kt}{S}\PYG{o}{]}    \PYG{k}{=} \PYG{o}{(}\PYG{n}{T} \PYG{k}{=\PYGZgt{}} \PYG{n}{S}\PYG{o}{)} \PYG{k}{=\PYGZgt{}} \PYG{n+nc}{Cancel}
\PYG{k}{type} \PYG{k+kt}{Publish}\PYG{o}{[}\PYG{k+kt}{T}, \PYG{k+kt}{S}\PYG{o}{]}      \PYG{k}{=} \PYG{n}{T} \PYG{k}{=\PYGZgt{}} \PYG{n+nc}{List}\PYG{o}{[}\PYG{k+kt}{S}\PYG{o}{]}
\end{Verbatim}

\code{T} is the type of the communicated item, and \code{S} is an optional type
(default \code{()}, called \code{void} in some languages) that callbacks may want
to pass back to the producer, to signal some sort of ``status''. The return
value of \code{Cancel} is whether the subscription was not already cancelled.

Even if absent from the language, having an idea on what types \emph{ought} to be
helps us to write combinators, e.g. to make a complex subscribe function
(``run \code{A} after event \code{X} but run \code{B} instead if event \code{Y} happens
first and run \code{A2} if event \code{X} happens after that'') or a complex cancel
function (``cancel all in set \code{X} and if all of them were already cancelled
then also cancel all in set \code{Y}'').

\item[{\code{EventContext}}] \leavevmode
A utility that supports efficient prefix-matched subscriptions, so consumers
can specify a filter for the items they're interested in. The type signature
of its public part is something like \code{\_Prefix\_{[}T{]} =\textgreater{} Subscribe{[}T, S{]}},
pretending for now that \code{\_Prefix\_} is a real type.

\item[{\code{Timer}}] \leavevmode
Execute something in the future. Its type is simply \code{Subscribe{[}Time, Unit{]}}
so that it can be used with combinators. When integrating our library into an
application, one can simply write an adapter that satisfies this interface,
for whichever execution framework is used.

\item[{\code{Future}}] \leavevmode
We only use these for user-level actions, so we don't need many combinators
for them. Standard libraries are adequate for our use cases, e.g. \code{Promise}
(JS) or \code{defer.Deferred} (Python).

\end{description}

We also have more complex utilities such as \code{Monitor}, built on top of
\code{Observable} and its friends, used to implement liveness and freshness
behaviours. For more details, see the API documentation \phantomsection\label{engineering:id4}{\hyperref[references:mpenc\string-api]{\crossref{{[}mpenc-api{]}}}}.

\subsection{High-level}
\label{engineering:high-level}\label{engineering:id7}
We define two interfaces (\emph{trait} or \emph{typeclass} in some languages) as a common
pattern for our \href{https://en.wikipedia.org/wiki/Actor\_model}{Actor}-like
components to follow. Each interface is a (function, subscribe-function) pair.
The former is to provide input into the component, the latter to accept output
from it.

One interface is for interacting with a more ``high level'' component, e.g. a
user interface:

\begin{Verbatim}[commandchars=\\\{\}]
\PYG{k}{trait} \PYG{n+nc}{ReceivingSender}\PYG{o}{[}\PYG{k+kt}{SendInput}, \PYG{k+kt}{RecvOutput}\PYG{o}{]} \PYG{o}{\PYGZob{}}
  \PYG{k}{def} \PYG{n}{send}   \PYG{k}{:} \PYG{k+kt}{SendInput} \PYG{o}{=\PYGZgt{}} \PYG{n+nc}{Boolean}
  \PYG{k}{def} \PYG{n}{onRecv} \PYG{k}{:} \PYG{k+kt}{Subscribe}\PYG{o}{[}\PYG{k+kt}{RecvOutput}, \PYG{k+kt}{Boolean}\PYG{o}{]}
    \PYG{c+c1}{// i.e. (RecvOutput =\PYGZgt{} Boolean) =\PYGZgt{} (() =\PYGZgt{} Boolean)}
\PYG{o}{\PYGZcb{}}
\end{Verbatim}

For example, when the UI wants to send some things to our session, it passes
this request to \code{Session.send}. To display things received from the session,
it hooks into \code{Session.onRecv}.

Another interface is for interacting with a more ``low level'' component, e.g. a
transport client:

\begin{Verbatim}[commandchars=\\\{\}]
\PYG{k}{trait} \PYG{n+nc}{SendingReceiver}\PYG{o}{[}\PYG{k+kt}{RecvInput}, \PYG{k+kt}{SendOutput}\PYG{o}{]} \PYG{o}{\PYGZob{}}
  \PYG{k}{def} \PYG{n}{recv}   \PYG{k}{:} \PYG{k+kt}{RecvInput} \PYG{o}{=\PYGZgt{}} \PYG{n+nc}{Boolean}
  \PYG{k}{def} \PYG{n}{onSend} \PYG{k}{:} \PYG{k+kt}{Subscribe}\PYG{o}{[}\PYG{k+kt}{SendOutput}, \PYG{k+kt}{Boolean}\PYG{o}{]}
    \PYG{c+c1}{// i.e. (SendOutput =\PYGZgt{} Boolean) =\PYGZgt{} (() =\PYGZgt{} Boolean)}
\PYG{o}{\PYGZcb{}}
\end{Verbatim}

For example, when we want to tell a GKA session membership operation that we
received a packet for it, we call \code{Greeting.recv}. To service its requests to
send out response packets, we hooks into \code{Greeting.onSend}.

Here are some examples of our components that implement the above interfaces:

\begin{Verbatim}[commandchars=\\\{\}]
\PYG{k}{trait} \PYG{n+nc}{Session}         \PYG{k}{extends} \PYG{n+nc}{ReceivingSender}\PYG{o}{[}\PYG{k+kt}{SessionAction}, \PYG{k+kt}{SessionNotice}\PYG{o}{]}\PYG{o}{;}
\PYG{k}{trait} \PYG{n+nc}{GroupChannel}    \PYG{k}{extends} \PYG{n+nc}{ReceivingSender}\PYG{o}{[}\PYG{k+kt}{ChannelAction}, \PYG{k+kt}{ChannelNotice}\PYG{o}{]}\PYG{o}{;}
\PYG{k}{trait} \PYG{n+nc}{Greeting}        \PYG{k}{extends} \PYG{n+nc}{SendingReceiver}\PYG{o}{[}\PYG{k+kt}{RawByteInput}, \PYG{k+kt}{RawByteOutput}\PYG{o}{]}\PYG{o}{;}
\PYG{k}{class} \PYG{n+nc}{SessionBase}     \PYG{k}{extends} \PYG{n+nc}{SendingReceiver}\PYG{o}{[}\PYG{k+kt}{RawByteInput}, \PYG{k+kt}{RawByteOutput}\PYG{o}{]}\PYG{o}{;}

\PYG{k}{type} \PYG{k+kt}{RawByteInput}     \PYG{o}{=} \PYG{o}{(}\PYG{n+nc}{SenderAddr}\PYG{o}{,} \PYG{n+nc}{Array}\PYG{o}{[}\PYG{k+kt}{Byte}\PYG{o}{]}\PYG{o}{)}
\PYG{k}{type} \PYG{k+kt}{RawByteOutput}    \PYG{o}{=} \PYG{o}{(}\PYG{n+nc}{Set}\PYG{o}{[}\PYG{k+kt}{RecipientAddr}\PYG{o}{]}\PYG{o}{,} \PYG{n+nc}{Array}\PYG{o}{[}\PYG{k+kt}{Byte}\PYG{o}{]}\PYG{o}{)}
\end{Verbatim}

These interfaces are also used privately too, to maintain a common style for
the code architecture. For example \code{HybridSession} contains an implementation
of \code{SendingReceiver{[}ChannelNotice, ChannelAction{]}}, but this is not exposed
since it is just an implementation detail, and it is only meant to be linked
with the associated \code{GroupChannel}.

We define \code{S} for \code{Subscribe{[}T, S{]}} as \code{Boolean} in these interfaces for
simplicity, meaning ``the item was \{accepted, rejected\} by the consumer''. This
allows us to detect errors -- such as transport failures in sending messages,
or trial decryption failures in receiving packets -- but in a loosely-coupled
way that discourages violation of the separation of layers. One reasonable
extension for the future, is to use a 3-value logic to represent \{accept, try
later, reject\}, which helps both of the previous cases.

This concludes the overview of our reference implementation. All the code that
is not mentioned here, is a straightforward application of software engineering
principles or algorithm writing, as applied to our protocol design (previous
chapter) and software design (this chapter). For more details, see the API
documentation \phantomsection\label{engineering:id8}{\hyperref[references:mpenc\string-api]{\crossref{{[}mpenc-api{]}}}} and/or source code.

\chapter{Cryptography}
\label{crypto::doc}\label{crypto:cryptography}
Here, we document the specifics of how we use and implement cryptography in our
protocol, and describe the information contained in our packets.

This is \emph{not} an overview of our full protocol system. Beyond processing and
generating packets, other algorithms and data structures must be implemented to
ensure a coherent high-level understanding of the state of the session, and the
ability to actively react to wider concerns such as liveness and consistency of
session state. These latter topics are covered in {\hyperref[protocol::doc]{\crossref{\DUrole{doc}{Protocol}}}}.

\section{Packet overview}
\label{crypto:packet-overview}
Every packet in our protocol is a sequence of records, similar to HTTP headers.
Each record has a type and a value, and some records may appear multiple times.

There are two packet types: (a) packets part of a membership change operation,
that we sometimes also call \emph{greeting packets} or \emph{greeting messages}; and (b)
packets that represent a logical message written by members of the session.

Both packet types include the following common records, that occur either at
or near the start of the sequence, in order:
\begin{itemize}
\item {} 
\code{MESSAGE\_SIGNATURE} -- This is a cryptographic signature for the rest of
the packet. This is calculated differently for greeting and data packets,
defined in the respective sections below.

\item {} 
\code{PROTOCOL\_VERSION} -- This indicates the protocol version being used, as a
16-bit unsigned integer. Our current version is \code{0x01}.

\item {} 
\code{MESSAGE\_TYPE} -- Indicates the packet type, \code{GREETING} or \code{DATA}.

\end{itemize}

\section{Membership changes}
\label{crypto:membership-changes}
Our membership change protocol consists of two subprotocols, one to derive an
ephemeral encryption key and one to derive ephemeral signature keys.

More modern protocols with better security properties exist, but ours is a
simple approach using elementary cryptography only, which should be easier to
understand and review. It will be improved later, but for now is adequate for
our stated security goals, and was quick to implement.

The subprotocols are described in the following chapters; please read these
first. We introduce some terminology there, that is needed to understand the
rest of this document.

\subsection{Group Key Agreement (GKA)}
\label{crypto_gka:group-key-agreement-gka}\label{crypto_gka::doc}
The group key agreement is conducted according to CLIQUES \phantomsection\label{crypto_gka:id1}{\hyperref[references:cliq]{\crossref{{[}00CLIQ{]}}}}.  We modify
the protocol to use ECDH based on x25519 instead of classic DH, since it is
superior in all attributes -- faster to process, smaller keys to transport and
store, and existing libraries are simpler to interface with.  It negotiates a
shared group key for \(n\) members in a total of \(O(n)\) messages
sent, and re-negotiates (single-member) include, exclude or key refresh in
\(O(1)\) messages, assuming broadcasts are available.

The CLIQUES key agreement algorithm is based on the Group-Diffie-Hellman (GDH)
concept, in which each participant generates a private contributory key portion
for the session.  This will be computed into the (public) intermediate keys
passed between the members.  At the end of the process, a set of \(n\)
intermediate keys will be available to all, with each of them lacking the
private contribution of exactly \emph{one} participant.  Each participant uses that
one respective intermediate key missing their own contribution to compute the
\emph{same} shared group key.

Based on this GDH concept two related protocols are constructed: The initial
key agreement (IKA) and the auxiliary key agreement (AKA).  IKA is used for an
initial key agreement with no initial knowledge of intermediate keys of the
other parties.  AKA is a supplementary and simplified protocol to agree on an
new, changed group key once the IKA protocol has successfully terminated.  It
is executed on including or excluding participants as well as refreshing the
shared group key.

Besides their slightly different actions within the group key agreement (e.g.
initiator of upflow or of downflow, see below), no participant of the IKA or
AKA bear a special role.  All participants of the CLIQUES key agreement can be
considered equal in terms of the protocol functionality.

CLIQUES does \textbf{not} protect against active attackers.  However, in our use of
it, all messages are co-located with our Authenticated Signature Key Exchange
(ASKE) messages, which authenticates each full message in the whole protocol.
Therefore, an active attack is countered by the final {\hyperref[crypto_aske:aske\string-verification]{\crossref{\DUrole{std,std-ref}{ASKE verification}}}} step.

\subsubsection{Initial Key Agreement (IKA)}
\label{crypto_gka:initial-key-agreement-ika}
The initial key agreement follows the outline of the \phantomsection\label{crypto_gka:id2}{\hyperref[references:cliq]{\crossref{{[}00CLIQ{]}}}} IKA.1 protocol.
The protocol contains two phases: The \emph{upflow} phase and the \emph{downflow} phase.
In the upflow, each participant generates their private key contribution, and
mixes it into the elements of the (growing) chain of intermediate keys.  Once
everyone has participated in the upflow, the last participant will initiate the
downflow by broadcasting the final chain of intermediate keys to all, enabling
them to individually compute the group key.

\paragraph{Upflow}
\label{crypto_gka:upflow}
The first participant assembles an ordered list of all participants included in
the IKA, with themselves as the first element.  Then, it generates its private
Diffie-Hellman key \(x_1\) and computes the corresponding public key using
DH exponentation.  In the case of mpENC, this is x25519 scalar multiplication
with the base point \(G\), i.e. \(x_1G\).  Then it generates a list of
intermediate keys to pass onto the next particiant in the list.  The initial
portion of the list (excluding the final element) are intermediate keys lacking
its own contribution; for the first participant, this is simply \((G)\).
The last -- or \emph{cardinal} -- element is an intermediate key that contains \emph{all}
previous participants' contributions; for the first participant, this is
\((x_1G)\).

Successive participants receiving the upflow messages similarly generate their
own private contributions.  This contribution is used for computing new
intermediate keys using DH exponentation.  The previous cardinal generates two
items in the new list -- itself unaltered, inserted just before the new
cardinal key, and also the new cardinal key, which is the previous cardinal key
multiplied with their own contribution.

\textbf{Example:}

The following figure shows the sequence of upflow messages (\(u_i\)) sent
among four participants (\(p_i\)).

\includegraphics{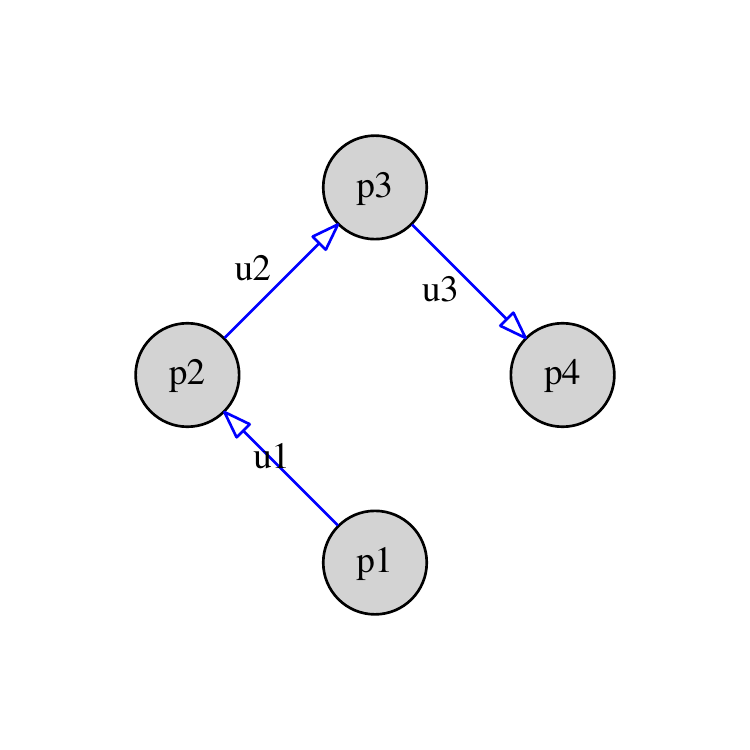}
\begin{description}
\item[{\(u1\) contains:}] \leavevmode\vspace{-\baselineskip}\begin{itemize}
\item {} 
Participants: \((p1,\; p2,\; p3,\; p4)\)

\item {} 
Calculate intermediate keys: \((1,\; x_1) . G\)

\item {} 
Intermediate keys: \((G,\; x_1G)\)

\end{itemize}

\item[{\(u2\) contains:}] \leavevmode\vspace{-\baselineskip}\begin{itemize}
\item {} 
Participants: \((p1,\; p2,\; p3,\; p4)\)

\item {} 
Calculate intermediate keys: \(x_2 . \mathsf{init}_1;\; (1,\; x_2) . \mathsf{ckey}_1\)

\item {} 
Intermediate keys: \((x_2G,\; x_1G,\; x_2x_1G)\)

\end{itemize}

\item[{\(u3\) contains:}] \leavevmode\vspace{-\baselineskip}\begin{itemize}
\item {} 
Participants: \((p1,\; p2,\; p3,\; p4)\)

\item {} 
Calculate intermediate keys: \(x_3 . \mathsf{init}_2;\; (1,\; x_3) . \mathsf{ckey}_2\)

\item {} 
Intermediate keys: \((x_3x_2G,\; x_3x_1G,\; x_2x_1G,\; x_3x_2x_1G)\)

\end{itemize}

\end{description}

Where \(A;\; B\) denotes vector concatenation, and \(\mathsf{init}_i\)
and \(\mathsf{ckey}_i\) denote respectively the initial portion and
cardinal key (final element) of the intermediate keys contained in \(u_i\).

\paragraph{Downflow}
\label{crypto_gka:downflow}
The last participant in the chain performs the same operations by adding their
own contributions to the intermediate keys.  However, the cardinal key at this
stage is complete, containing contributions from everyone.  Therefore, the last
participant retains the new cardinal key as the shared group secret, and
broadcasts the list of other intermediate keys to \emph{all} other members.  After
that, the participant has finished their participation in the IKA protocol and
possesses the shared group secret, then computes the group key from it.

Each recipient of the downflow message will now be able to take ``their''
intermediate key out of the list (i.e. the one missing their own contribution).
For the \(i\)-th member in the chain, this is the \(i\)-th intermediate
key.  Through DH exponentiation of their own private key contribution with
``their'' intermediate key, they will all derive the same shared secret.  This is
the point this participant has also completed its part in the IKA and has
transitioned into the ready state.

\textbf{Example:}

The following figure shows the corresponding downflow message (\(d\))
broadcast to all other participants (\(p_i\)).

\includegraphics{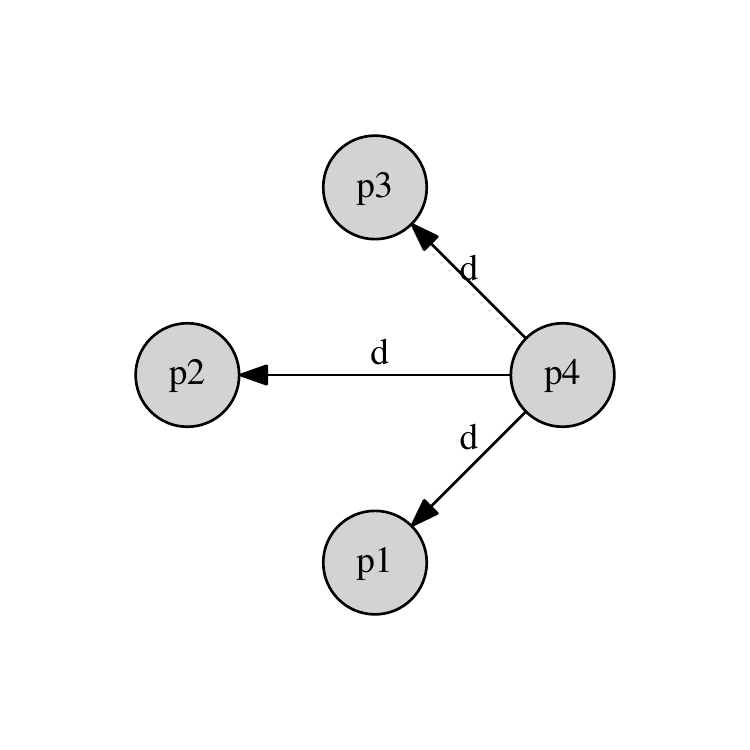}
\begin{description}
\item[{\(d\) contains:}] \leavevmode\vspace{-\baselineskip}\begin{itemize}
\item {} 
Participants: \((p1,\; p2,\; p3,\; p4)\)

\item {} 
Intermediate keys: \((x_4x_3x_2G,\; x_4x_3x_1G,\; x_4x_2x_1G,\; x_3x_2x_1G)\)

\end{itemize}

\end{description}

After receiving these intermediate keys, every participant can compute the same
shared group secret by multiplying ``their'' intermediate key with their own
private contribution:
\begin{equation*}
\begin{split}x_1x_4x_3x_2G = x_2x_4x_3x_1G = x_3x_4x_2x_1G = x_4x_3x_2x_1G\end{split}
\end{equation*}
This group secret is used as input into a KDF to derive further keys to be used
for other operations, such as message encryption to the group.

\subsubsection{Auxiliary Key Agreement (AKA)}
\label{crypto_gka:auxiliary-key-agreement-aka}
Once an initialised chat encryption is available for an established group of
participants, an auxiliary key agreement (AKA) can be invoked.  These runs are
necessary for changes in group participants (including new members or excluding
existing ones) to update the group secret.  Therefore allowing the previous
participant set only to read messages before the AKA, and the new participant
set to read/write messages after the AKA.  Furthermore the AKA can also be used
to refresh the group secret, for more fine-grained forward secrecy, by updating
a participant's private key contribution.

\paragraph{Member Inclusion}
\label{crypto_gka:member-inclusion}\label{crypto_gka:gka-member-include}
Member inclusion is performed very similarly to the IKA protocol.  An existing
participant may initiate an upflow for this.  First the new participant(s) are
appended to the list of existing participants.  To avoid the new participants
gaining knowledge of the previous group secret, the initiator of the include is
required to update its private key contribution in the following fashion:
\begin{enumerate}
\item {} 
Perform a DH exponentiation with its own private contribution on its ``own''
intermediate key (as if it was generating the old group secret), then append
it to the list of intermediate keys for each new member.  Note that this is
a secret value and must not be sent yet! The next steps hide it.

\item {} 
Generate a new private key contribution (see {\hyperref[crypto_gka:note\string-key\string-contributions]{\crossref{\DUrole{std,std-ref}{Updating Private Key Contributions}}}}).

\item {} 
Perform DH exponentiations on all intermediate keys, except its ``own'', with
the new private key contribution.

\end{enumerate}

The upflow is now initiated by sending this list of updated intermediate keys
to the (first of the) new participant(s) to include.  The new participant(s)
perform the key agreement protocol in exactly the same fashion as done in the
IKA upflow by generating their own private key contributions, performing DH
computations with them on the intermediate keys and extending the intermediate
key list with their ``own'' intermediate key.

The last (new) participant in the extended list now will initiate the downflow
broadcast message consisting of \emph{all} intermediate keys, thus enabling every
participant to compute the new shared group secret and reach a ready state.

Using the AKA for includes it is possible to add new participants either one by
one or multiple at the same time.  It is more efficient to add multiple new
participants at the same time than to add them sequentially.

\textbf{Example:}

The following figure shows inclusion of a participant (\(p5\)) -- initiated
by \(p1\) -- to the existing group of four participants.

\includegraphics{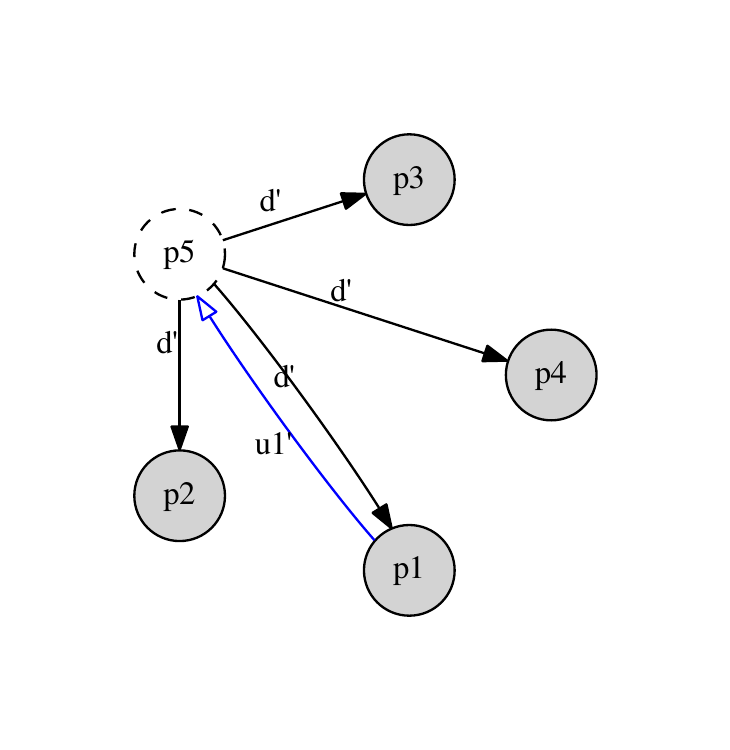}
\begin{description}
\item[{\(u1'\) contains:}] \leavevmode\vspace{-\baselineskip}\begin{itemize}
\item {} 
Participants: \((p1,\; p2,\; p3,\; p4,\; p5)\)

\item {} 
Intermediate keys: \((x_4x_3x_2G,\; x_1'x_4x_3x_1G,\;
x_1'x_4x_2x_1G,\; x_1'x_3x_2x_1G,\; x_1'x_1x_4x_3x_2G)\)

\end{itemize}

\item[{\(d'\) contains:}] \leavevmode\vspace{-\baselineskip}\begin{itemize}
\item {} 
Participants: \((p1,\; p2,\; p3,\; p4,\; p5)\)

\item {} 
Intermediate keys: \((x_5x_4x_3x_2G,\; x_5x_1'x_4x_3x_1G,\;
x_5x_1'x_4x_2x_1G,\; x_5x_1'x_3x_2x_1G,\; x_1'x_1x_4x_3x_2G)\)

\end{itemize}

\end{description}

Where \(x_1\) is the initiator's old private key contribution, \(x_1'\)
is the new contribution.

Again, after receiving these intermediate keys, every participant can compute
the same shared group secret by multiplying ``their'' intermediate key with their
own private contribution(s):
\begin{equation*}
\begin{split}x_1'x_1x_5x_4x_3x_2G = x_2x_5x_1'x_4x_3x_1G = x_3x_5x_1'x_4x_2x_1G =
x_4x_5x_1'x_3x_2x_1G = x_5x_1'x_1x_4x_3x_2G\end{split}
\end{equation*}

\paragraph{Member Exclusion}
\label{crypto_gka:member-exclusion}
The AKA protocol flow for member exclusion is similar to -- but simpler -- than
member inclusion.  The initiator updates their private key contribution (see
{\hyperref[crypto_gka:note\string-key\string-contributions]{\crossref{\DUrole{std,std-ref}{Updating Private Key Contributions}}}}) in the same manner as for includes above.  Then
the participant(s) as well as their intermediate key(s) are removed from the
respective lists for the participant(s) to be excluded.  Now the downflow
broadcast message can be sent directly without the need of a preceding upflow
phase.  Thus, all remaining participants can compute the new shared group
secret and reach a ready state.

When using the AKA for exclusion it is possible to remove participants either
one by one or multiple at the same time.  It is more efficient to remove
multiple participants at the same time than to remove them sequentially.

\paragraph{Key Refresh}
\label{crypto_gka:key-refresh}
To help more granular forward secrecy over extended periods of key use, it is a
good idea to refresh the group secret at suitable intervals (e.g. depending on
time, number of messages or volume encrypted with it).  A key refresh is very
simple, and can be initiated by \emph{any} participant.  The initiating participant
renews their own private key contribution (see {\hyperref[crypto_gka:note\string-key\string-contributions]{\crossref{\DUrole{std,std-ref}{Updating Private Key Contributions}}}}),
and broadcasts a downflow message with all updated intermediate keys to all
participants without the need of a preceding upflow.  Thus, all participants
can compute the new shared group key and reach a ready state.

It is wise for participants to track the ``age'' of their own private key
contribution.  This mechanism can be used for achieving a ``rolling'' group
secret refresh by always updating the oldest private key contributions of
participants.

\subsubsection{Member Departure}
\label{crypto_gka:member-departure}
Member departure is the voluntary parting of a participant rather than an
exclusion initiated by another participant.  In effect it is the same, with the
only difference that the departing member indicates the desire to leave, and a
member exclusion AKA will be initiated upon that by another participant.

In mpENC, this is not a direct concern of the GKA, and works the same way
independently of the particular GKA we choose.  That is, the ``desire to leave''
is a special data message, sent via the normal mechanism for data messages.

\subsubsection{Updating Private Key Contributions}
\label{crypto_gka:updating-private-key-contributions}\label{crypto_gka:note-key-contributions}
When the private key contribution (for an inclusion, exclusion or refresh) is
updated, the client must keep \emph{all} the key contributions in a list, including
old contributions.  When performing computations to derive a new cardinal key,
this whole list of one's own private key contributions needs to be used.

In theory, these individual contributions can be condensed into a single value,
via multiplication modulo the order of the base element (base point in ECC).
However, in x25519 only certain values are valid secret keys; secret inputs not
in the expected format are coerced \footnote[1]{\sphinxAtStartFootnote%
For example, in \href{https://github.com/jedisct1/libsodium/blob/6aacecac/src/libsodium/crypto\_scalarmult/curve25519/ref10/scalarmult\_curve25519\_ref10.c\#L26}{libsodium}
and \href{https://github.com/meganz/jodid25519/blob/d9857d48/src/jodid25519/curve255.js\#L83}{jodid25519}.
} into this format, which effectively
changes the value used for the actual mathematical scalar multiplication.  If
we combine secret keys using modular multiplication, this will sometimes result
in a value that is effectively corrupted by typical x25519 APIs.  So, we cannot
do this in practise; we must store all our contributions separately, to be
mixed individually into our intermediate key later.

This sequence may grow big over time, so that the overhead of applying a long
sequence of elliptic curve scalar multiplications can become more significant.
In such cases, it may be worth to re-key the whole session.  We have not yet
implemented this, but will do so if it becomes a problem in practice.

Additionally, we cannot pre-emptively combine old contributions into the
intermediate key, e.g. to add an extra step in our key-update sequence
described in {\hyperref[crypto_gka:gka\string-member\string-include]{\crossref{\DUrole{std,std-ref}{Member Inclusion}}}}:
\begin{enumerate}
\setcounter{enumi}{3}
\item {} 
Perform DH exponentiations on its ``own'' intermediate key, with the \emph{old}
private key contribution (as from step \#1).

\end{enumerate}

This would cause us to reveal the group secret of the previous session, namely
\(x_1x_5x_4x_3x_2G\) in the example of the above section, which of course
would be a catastrophic security failure.

\subsection{Authenticated Signature Key Exchange (ASKE)}
\label{crypto_aske::doc}\label{crypto_aske:authenticated-signature-key-exchange-aske}
As previously noted, we must authenticate the user's membership in the session,
as well as authenticate the content of their messages.  This could easily be
accomplished by using static private keys to make signatures.  However, this
destroys any chance that we might have in the future at retaining deniability
(also known as repudiability) of ciphertext.  Once this property is lost, we
can never regain it in a higher layer, and it is critical for \emph{confidentiality
of metadata}, so it is better to retain it here.

Instead, we generate ed25519 ephemeral signing keys for each participant, to be
used to authenticate messages for the current session only.  We also derive a
session ID from participants' nonces to ensure freshness.  At the end of the
session the ephemeral private signing key may be published to the transport,
allowing for retrospective chat transcript alteration, and therefore allowing
repudiation of the contents of transcripts presented post-session.  However,
our mechanism \emph{does not} provide deniability of the \emph{participation} in the
authenticated signature key exchange (ASKE) protocol.

Our approach takes some loose inspiration from \phantomsection\label{crypto_aske:id1}{\hyperref[references:dgka]{\crossref{{[}06DGKA{]}}}} and \phantomsection\label{crypto_aske:id2}{\hyperref[references:dske]{\crossref{{[}13DSKE{]}}}}, and may
also be compared with \phantomsection\label{crypto_aske:id3}{\hyperref[references:spag]{\crossref{{[}03SPAG{]}}}} and \phantomsection\label{crypto_aske:id4}{\hyperref[references:sdgk]{\crossref{{[}12SDGK{]}}}}.  Both are constructions to turn
an unauthenticated GKA into an authenticated one.  The former construction is
not deniable, whereas the latter is fully deniable (including participation in
the session) but requires more advanced cryptography (ring signatures) that
have not yet seen widespread usage.

Our {\hyperref[crypto_gka::doc]{\crossref{\DUrole{doc}{GKA}}}} protocol consists of an upflow (sequential collect)
and downflow (broadcast) phase, whereas our ASKE protocol is a constant-round
broadcast protocol, as are the other authenticated agreements referenced above.
To make our {\hyperref[crypto::doc]{\crossref{\DUrole{doc}{combined protocol}}}} easier to understand, the first
broadcast round has been serialised into a ``collection phase'' upflow.  The
first broadcast of the last member in the chain is the start of the downflow,
which is followed by an acknowledgement broadcast by every other participant.
Note that these latter broadcasts are not present in our group key agreement.

ASKE consists of three phases.  First each participant generates a nonce and an
ephemeral signature key pair, and forwards the nonce and public key (upflow).
In the second phase -- when in possession of all nonces -- each member
independently computes the shared session ID, and authenticates their ephemeral
signing key and session ID using their static private key (downflow).  Lastly,
each participant verifies all received acknowledgement messages.

\subsubsection{Initial Protocol Run}
\label{crypto_aske:initial-protocol-run}

\paragraph{Phase 1 -- Collection}
\label{crypto_aske:phase-1-collection}
In the first phase the session initiator \(i\) with the participant ID
\(\mathsf{pid}_i\) compiles an ordered list of all group members (their
participant IDs).  Additionally an empty list for the all participants' nonces
and ephemeral public keys is initialised.

The initiator then generates a nonce \(k_i\) and an ephemeral signature key
pair \((e_i, E_i)\).  They add these to the nonces and public keys lists.
They then send the participants' list (\(\mathsf{pid}_i\)), nonces list
(\(k_i\)) and public keys list (\(E_i\)) on to the next member in the
list, who again generates a nonce and ephemeral signature key pair to send on.

This phase ends with the last member in the list to add their contributions.
This last member is the initiator of the second phase.

\textbf{Example:}

The following figure shows the sequence of upflow messages
(\(u_i\)) sent among four participants (\(p_i\)).

\includegraphics{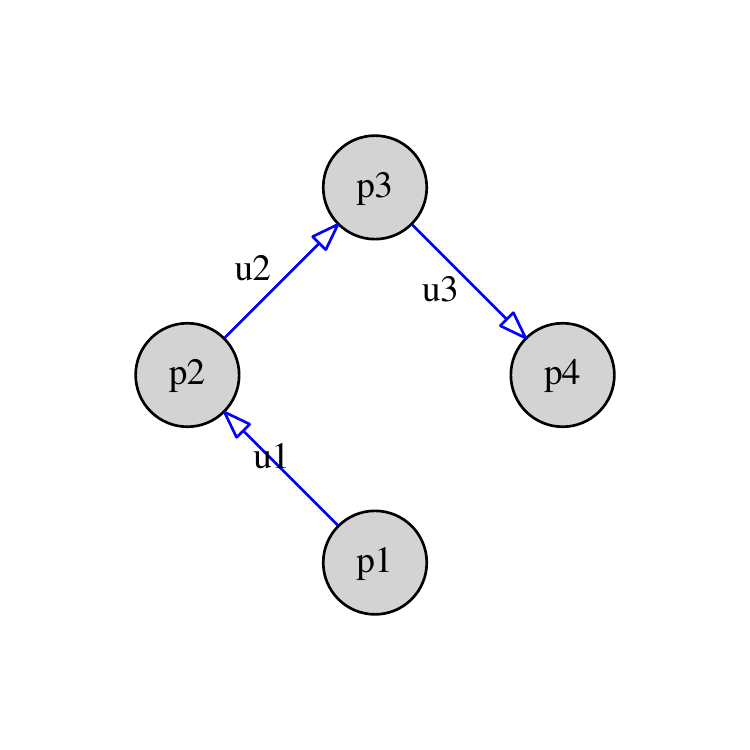}
\begin{description}
\item[{\(u1\) contains:}] \leavevmode\vspace{-\baselineskip}\begin{itemize}
\item {} 
Participants: \((p1,\; p2,\; p3,\; p4)\)

\item {} 
Nonces: \((k_1)\)

\item {} 
Ephemeral public signing keys: \((E_1)\)

\end{itemize}

\item[{\(u2\) contains:}] \leavevmode\vspace{-\baselineskip}\begin{itemize}
\item {} 
Participants: \((p1,\; p2,\; p3,\; p4)\)

\item {} 
Nonces: \((k_1,\; k_2)\)

\item {} 
Ephemeral public signing keys: \((E_1,\; E_2)\)

\end{itemize}

\item[{\(u3\) contains:}] \leavevmode\vspace{-\baselineskip}\begin{itemize}
\item {} 
Participants: \((p1,\; p2,\; p3,\; p4)\)

\item {} 
Nonces: \((k_1,\; k_2,\; k_3)\)

\item {} 
Ephemeral public signing keys: \((E_1,\; E_2,\; E_3)\)

\end{itemize}

\end{description}

\paragraph{Phase 2 -- Acknowledgement}
\label{crypto_aske:phase-2-acknowledgement}\label{crypto_aske:aske-session-sig}
The initiator of the downflow in the acknowledgement phase first constructs an
authenticator message from their own contributions:
\begin{equation*}
\begin{split}m_i = (\mathsf{CTX}||\mathsf{pid}_i||E_i||k_i||\mathsf{sid})\end{split}
\end{equation*}
Here, \(\mathsf{CTX}\) is a fixed byte sequence to prevent the signature
being used in another application; for this protocol version we use \code{acksig}.
\(\mathsf{sid}\) is the \emph{session ID}, calculated from all participant IDs
and nonces using a hash function \(H\):
\begin{equation*}
\begin{split}\mathsf{sid} = H(\mathsf{pid}_1||\mathsf{pid}_2||\ldots||\mathsf{pid}_n||k_1||k_2||\ldots||k_n)\end{split}
\end{equation*}
The IDs and nonces must be strictly ordered.  For mpENC on the Mega platform
the participant IDs are the full XMPP JIDs, and sorting is performed in lexical
order.  The nonces are ordered so as to correspond to their participant IDs.

Then, the initiator broadcasts the first message in the downflow, containing
the now-completed lists of participants (\(\mathsf{pid}_i\)), nonces
(\(k_i\)) and public keys (\(E_i\)), for all \(i\), along with a
signature of their own authenticator message \(\sigma_{s_i}(m_i)\) computed
with the static identity signature key \((s_i, S_i)\).  The purpose of this
is to authenticate all information contributed by the signing participant, as
well as what they believe the contributions of all session members to be.
\footnote[1]{\sphinxAtStartFootnote%
Although \(k_i\) is already ``contained in'' the session ID, we
explicitly add it to \(m_i\), to avoid depending on the security of its
calculation. This is hoped to simplify any future formal analysis.
}  Note that the authenticator message itself needs not be, and is not,
broadcast.

After receiving this, every participant is in possession of the information
required to calculate the supposed \(\mathsf{sid}\) for themselves, produce
what each \(m_i\) \emph{should be} and verify the \(\sigma_{s_i}(m_i)\) that
it \emph{should have} based on this information.

Now, each participant computes the session ID (\(\mathsf{sid}\)) from the
content of this initial broadcast message, checking that the values supposedly
contributed by them actually match what they output during the upflow phase.
Then, they generate their own authenticator message, corresponding signature,
and broadcast this signature to others.  The lists of intermediate values are
not necessary in these further broadcasts.

\textbf{Example:}

The following figure shows the corresponding downflow message
(\(d4\)) broadcast to all participants by \(p4\).

\includegraphics{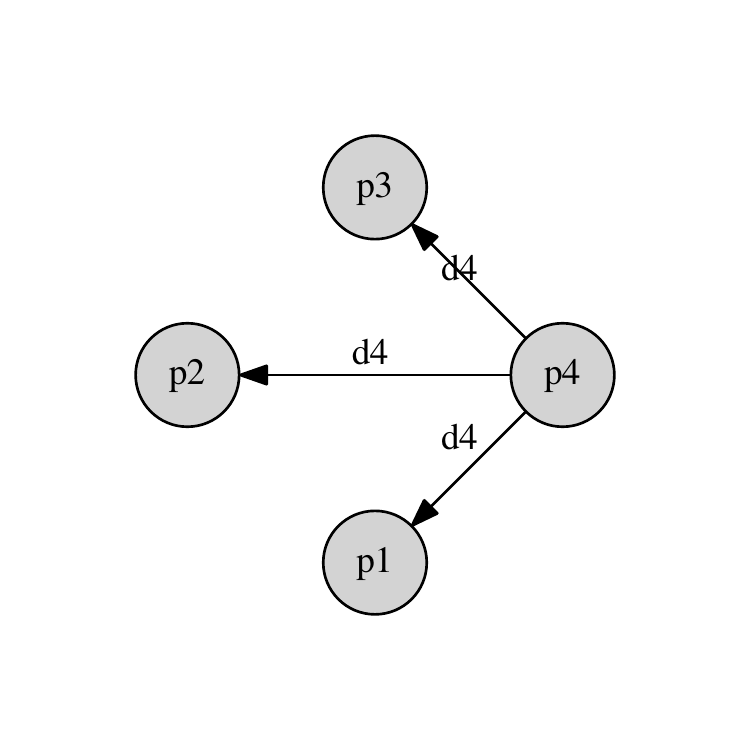}
\begin{description}
\item[{\(d4\) contains:}] \leavevmode\vspace{-\baselineskip}\begin{itemize}
\item {} 
Participants: \((p1,\; p2,\; p3,\; p4)\)

\item {} 
Nonces: \((k_1,\; k_2,\; k_3,\; k_4)\)

\item {} 
Ephemeral public signing keys: \((E_1,\; E_2,\; E_3,\; E_4)\)

\item {} 
Session signature: \(\sigma_{s_4}(m_4)\)

\end{itemize}

\end{description}

Upon receipt of \(d4\) every other participant sends out an
analogous \(dX\) message including their \emph{own} session signature.

\paragraph{Phase 3 -- Verification}
\label{crypto_aske:aske-verification}\label{crypto_aske:phase-3-verification}
This last phase does not require further messages to be sent.  Each participant
verifies the content of each received acknowledgement broadcast message against
their own available information.  The purpose is to have the assurance that all
participants are actively participating (avoids replays) with a fresh session,
and to have the assurance that the session's ephemeral signing keys are really
from the users that one is communicating with.

More specifically, as each participant receives each subsequent downflow
broadcast from \(\mathsf{pid}_i\), they compute \(m_i\) from the same
information used to compute their local value for \(\mathsf{sid}\), and
verify the signature contained in the received message (which is supposed to be
\(\sigma_{s_i}(m_i)\)) using the sender's long term static key \(S_i\).

The protocol completes successfully when all session signatures from all other
participants have been successfully verified against the local session ID and
each participant's static identity signature key.

Following successful completion, \emph{only} the ephemeral keys are needed for
message authentication -- signing with the static keys would effectively
inhibit any plausible deniability.  However the static keys are needed for
further changes to the session membership.

\subsubsection{Auxiliary Protocol Runs}
\label{crypto_aske:auxiliary-protocol-runs}
Upon changing the participant composition of the chat (inclusions or exclusions
of members) some session information changes: The list of participants, nonces
and ephemeral signing keys.  Therefore, the session ID also changes.

\paragraph{Member Inclusion}
\label{crypto_aske:member-inclusion}
To include participants, the initiator extends the list of participants by the
new participant(s).  A new collection (upflow) message is sent to the (first)
new participant, including the \emph{new} list of participants \(p_i\) and
\emph{already existing} nonces \(k_i\) and ephemeral signing keys \(E_i\).
The collection upflow percolates through all new participants, and the last one
will initiate a new acknowledgement downflow phase followed by a verification
phase identically to the initial protocol flow as outlined above.

\textbf{Example:}

The following figure shows addition of a participant (\(p5\)) -- initiated
by \(p1\) -- to the existing group of four participants.

\includegraphics{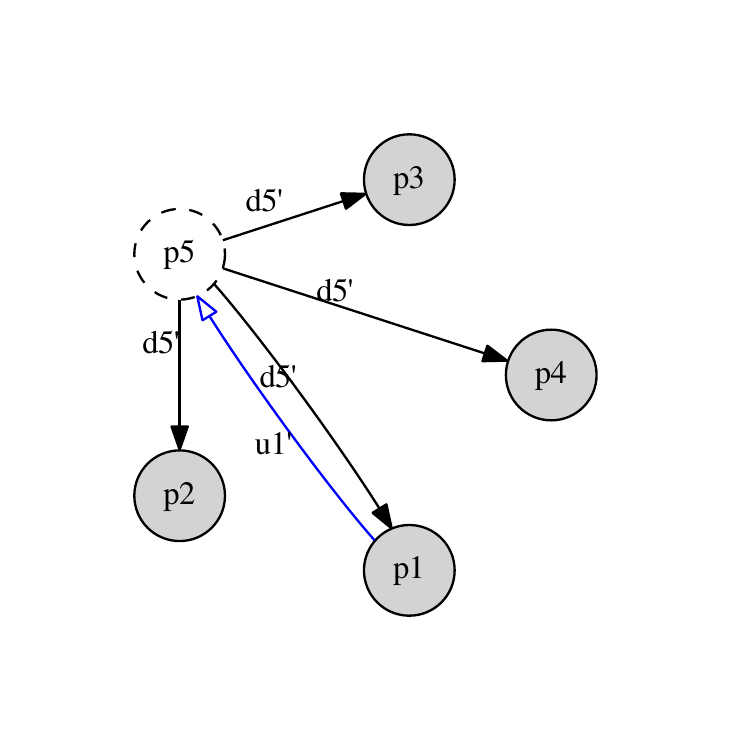}
\begin{description}
\item[{\(u1'\) contains:}] \leavevmode\vspace{-\baselineskip}\begin{itemize}
\item {} 
Participants: \((p1,\; p2,\; p3,\; p4,\; p5)\)

\item {} 
Nonces: \((k_1,\; k_2,\; k_3,\; k_4)\)

\item {} 
Ephemeral public signing keys: \((E_1,\; E_2,\; E_3,\; E_4)\)

\end{itemize}

\item[{\(d5'\) contains:}] \leavevmode\vspace{-\baselineskip}\begin{itemize}
\item {} 
Participants: \((p1,\; p2,\; p3,\; p4,\; p5)\)

\item {} 
Nonces: \((k_1,\; k_2,\; k_3,\; k_4,\; k_5)\)

\item {} 
Ephemeral public signing keys: \((E_1,\; E_2,\; E_3,\; E_4,\; E_5)\)

\item {} 
Session signature: \(\sigma_{s_5}(m_5)\)

\end{itemize}

\end{description}

After receiving this message, \(p1\) through \(p4\) will likewise
broadcast their acknowledgement messages to all participants as well as verify
all received session signatures \(\sigma_{s_i}(m_i)\).

\paragraph{Member Exclusion}
\label{crypto_aske:member-exclusion}
On member exclusion, the process is simpler as it does not require a collection
(upflow) phase, as all remaining participants have announced already.  The
initiator of the exclusion removes the excluded participant(s) from the list of
participants, and their respective nonces and ephemeral signing keys are as
well removed.

Importantly, the initiator \emph{must} update their own nonce to prevent collisions
in the session ID \(\mathsf{sid}\) with a previous session ID consisting of
the same set of participants.  They then compute a new session ID and session
signature \(\sigma_{s_i}(m_i)\) from these updated values, and used them to
directly broadcast the initial downflow message to all remaining participants.
Each of them again verifies all session signatures and broadcasts their own
acknowledgement (if still outstanding).

\paragraph{Key Refresh}
\label{crypto_aske:key-refresh}
The concept of a key refresh for ASKE is currently not considered.

\subsubsection{Member Departure}
\label{crypto_aske:member-departure}
As with our GKA, our ASKE does not include a \emph{member departure} operation, this
instead being handled in a different part of the wider protocol.

In the future, our departure mechanism will include publishing the ephemeral
signature key, to support a limited form of ciphertext deniability.  This is
described more in {\hyperref[future\string-work:publish\string-sess\string-sig\string-keys]{\crossref{\DUrole{std,std-ref}{Publish signature keys}}}}.

\subsubsection{Confirmation of the Shared Group Key}
\label{crypto_aske:confirmation-of-the-shared-group-key}
The ASKE mentioned above only protects the GKA against external attackers.  A
malicious insider can cause different participants to generate different shared
group keys.  We \emph{could} protect against this by adding the shared group secret
\(x_1 \dots x_nG\) (or better, something derived from it) to the definition
of \(m_i\). However, we currently omit this, for the following reasons:
\begin{itemize}
\item {} 
This would cause a key refresh to require everyone's acknowledgement, adding
an extra round to it.

\item {} 
This attack does no more damage beyond ``drop all messages'', which is already
available to active transport attackers:
\begin{itemize}
\item {} 
For authenticity of session membership, i.e. \emph{entity authentication} or
\emph{freshness}, all participants are indeed part of the session, having
participated in the ASKE and verified each others' session signatures. They
just may have generated different encryption keys.

\item {} 
For authenticity of message content: each ephemeral signature includes a
hash of something derived from the group key, so one will never correctly
verify-decrypt a message that was encrypted using a different group key
from yours.

\item {} 
For authenticity of message membership: our other checks (for reliability
and consistency) will cause each side of a ``split'' to timeout and emit
security warnings, because the other side was unable to acknowledge them.

\end{itemize}

\end{itemize}

However, it is not too much complexity to add such a feature, if the attack
turns out to be a real problem.  Another benefit of adding the protection would
be to make the above argument unnecessary, which reduces the complexity cost of
analysing our protocol.

\subsection{Combined authenticated group key agreement}
\label{crypto:combined-authenticated-group-key-agreement}
There are four operation types: \emph{establish session}, \emph{include member}, \emph{exclude
member} and \emph{refresh group key}. Refresh may be thought of as a ``no-op'' member
change and that is how our overall protocol system treats it. Each operation
consists of the following stages:
\begin{description}
\item[{Protocol upflow (optional)}] \leavevmode
The protocol upflow consists of a series of directed messages; one user sends
it to exactly one recipient, i.e. the next member in the \code{MEMBER} list of
records. It is used to ``collect'' information from one member to the next. In
a transport that \emph{only} offers broadcast, the other recipients simply ignore
packets not meant for them (i.e. the \code{DEST} record is not them).

Every upflow message includes the records marked {[}1{]} from the summary below.
The first upflow message also includes records marked {[}2{]}, which contains
metadata about the operation, used by the concurrency resolver.

This stage is not present for exclude or refresh operations.

\item[{Protocol downflow}] \leavevmode
The protocol downflow consists of a series of broadcast messages, sent to all
recipients in the group. It is used to ``distribute'' information contributed
by all participants (or derived therefrom) to all participants at once.

Every downflow message includes the records marked {[}c{]} from the summary
below. The first downflow message also includes records marked {[}a{]}, which
contains information from everyone, collected during the upflow. If the
operation has no upflow (e.g. exclude and refresh), the first downflow
message also includes records marked {[}b{]}.

\end{description}

A greeting packet contains the following records, in order:
\begin{itemize}
\item {} 
\code{MESSAGE\_SIGNATURE}, \code{PROTOCOL\_VERSION}, \code{MESSAGE\_TYPE} -- as above.

\item {} 
\code{GREET\_TYPE} -- What operation the packet is part of, as well as which
stage of that operation the packet belongs to. \footnote[1]{\sphinxAtStartFootnote%
The exact details may be viewed in the source code. The current
values are vestiges from previous iterations of the protocol, where we did
things differently and with more variation. A better approach would be to
infer this from the other records that are already part of the message, and
eliminate the redundant information, which may be set to incorrect values
by malicious participants. This will be done in a later iteration.
}

\item {} 
\code{SOURCE} -- User ID of the packet originator.

\item {} 
\code{DEST} (optional) -- User ID of the packet destination; if omitted, it
means this is a broadcast (i.e. downflow) packet.

\item {} 
\code{MEMBER} (multiple) -- Participating member user IDs. This record appears
\(n\) times, once for each member participating in this operation, i.e.
the set of members in the subsession that would be created on its success.
This also defines the orders of participants in the upflow sequence.

\item {} 
{[}a{]} \code{INT\_KEY} (multiple) -- Intermediate DH values for the GKA.

\item {} 
{[}a{]} \code{NONCE} (multiple) -- Nonces of each member for the ASKE.

\item {} 
{[}a{]} \code{PUB\_KEY} (multiple) -- Ephemeral session public signing keys of each
member.

\item {} 
{[}b{]} \code{PREV\_PF} -- Packet ID of the final packet of the previously completed
operation, or a random string if this is the first operation.

\item {} 
{[}b{]} \code{CHAIN\_HASH} -- Chain hash corresponding to that packet, as calculated
by the initiator of this operation. This allows joining members to calculate
subsequent hashes of their own, to compare with others for consistency.

\item {} 
{[}b{]} \code{LATEST\_PM} (multiple) -- References to the latest (logical, i.e. data)
messages from the previous subsession that the initiator of this operation
had seen, from when they initiated the operation.

\item {} 
{[}c{]} \code{SESSION\_SIGNATURE} -- Authenticated {\hyperref[crypto_aske:aske\string-session\string-sig]{\crossref{\DUrole{std,std-ref}{confirmation signature}}}} for the ASKE, signed with the long-term identity key of
the packet author.

\end{itemize}
\begin{enumerate}
\item {} 
Only for upflow messages and the first downflow message. During the upflow,
this gets filled to a maximum of \(n\) occurences.

\item {} 
Only for the initial packet of any operation. These fields are ignored by
the greeting protocol; instead they are used by the concurrency resolver.
See {\hyperref[protocol:transport\string-integration]{\crossref{\DUrole{std,std-ref}{Transport integration}}}} and \phantomsection\label{crypto:id2}{\hyperref[references:msa\string-5h0]{\crossref{{[}msa-5h0{]}}}} for details.

\item {} 
Only for downflow messages, to confirm the ASKE keys.

\end{enumerate}

\code{MESSAGE\_SIGNATURE} is made by the ephemeral signing key and signs the byte
sequence \(\mathsf{CTX} || \mathsf{S}\). \(\mathsf{CTX}\) is a fixed
byte sequence to prevent the signature being injected elsewhere; for greeting
packets in this version, it is \code{greetmsgsig}. \(\mathsf{S}\) is the byte
sequence of all subsequent records, starting with \code{PROTOCOL\_VERSION}.

Even though this ephemeral key is not authenticated against any identity keys
at the start of the operation, it is so by the end via \code{SESSION\_SIGNATURE}.
It is important not to allow ongoing operations to affect anything else in the
overall session, until this authentication has taken place; our atomic design
satisfies this security requirement.

Likewise, members may corrupt the structure of packets as the operation runs,
such as re-ordering the \code{MEMBER} list of records. However, there should be no
security risk; inconsistent results will be detected via \code{SESSION\_SIGNATURE}
and cause a failure of the overall operation.

\section{Messages}
\label{crypto:messages}
Once a subsession has been set-up after the completion of a membership change,
members may exchange messages confidentially (using the shared group key) and
authentically (using the ephemeral signing key).

A data packet contains the following records, in order:
\begin{itemize}
\item {} 
\code{SIDKEY\_HINT} -- A one byte hint, that securely gives the recipient an aid
in efficiently selecting the decryption key for this message.

\item {} 
\code{MESSAGE\_SIGNATURE}, \code{PROTOCOL\_VERSION}, \code{MESSAGE\_TYPE} -- as above.

\item {} 
\code{MESSAGE\_IV} -- Initialisation vector for the symmetric block cipher. The
cipher we choose is malleable, to give better deniability when we publish
ephemeral signature keys, similar to OTR \phantomsection\label{crypto:id3}{\hyperref[references:otr\string-spec]{\crossref{{[}OTR-spec{]}}}}.

\item {} 
\code{MESSAGE\_PAYLOAD} -- Ciphertext payload. The plaintext is itself a sequence
of records, as follows:
\begin{itemize}
\item {} 
\code{MESSAGE\_PARENT} (multiple) -- References to the latest (logical, i.e.
data) messages that the author had seen, when they wrote the message.

\item {} 
\code{MESSAGE\_BODY} -- UTF-8 encoded payload, the message content as written
by the human author.

\end{itemize}

\end{itemize}

\code{MESSAGE\_SIGNATURE} is made by the ephemeral signing key and signs the byte
sequence \(\mathsf{CTX} || H(\mathsf{sid} || \mathsf{gk}) || \mathsf{S}\).
\(\mathsf{CTX}\) is a fixed byte sequence to prevent the signature being
injected elsewhere; for data packets in this version, it is \code{datamsgsig}.
\(\mathsf{S}\) is the byte sequence of all subsequent records, starting
with \code{PROTOCOL\_VERSION}.

Since a single ephemeral signing key may be used across several subsessions,
the \(H(\dots)\) value ensures that each ciphertext is verifiably bound to
a specific subsession. \footnote[2]{\sphinxAtStartFootnote%
When/if we come to publish ephemeral signature keys, we will also
have to publish all \(H(\mathsf{sid} || \mathsf{gk})\) values
that were used by the key, to ensure that a forger can generate valid
packets without knowing the group encryption keys.
} \(\mathsf{sid}\) is the session ID (of the
subsession) as determined from the ASKE, and \(\mathsf{gk}\) is the
encryption key derived from the GKA.

\textbf{Padding}. We use an opportunistic padding scheme that obfuscates the lower
bits of the length of the message. This has not been analysed under an
aggressive threat model, but is very simple and cannot possibly be harmful. The
scheme is as follows:
\begin{itemize}
\item {} 
Define a baseline size \code{size\_bl}. We have chosen 128 bytes for this, to
accommodate the majority (e.g. see \phantomsection\label{crypto:id5}{\hyperref[references:mcic]{\crossref{{[}08MCIC{]}}}}) of messages in chats, while
still remaining a multiple of our chosen cipher's block size of 16 bytes.

\item {} 
The value of \code{MESSAGE\_BODY} is prepended by a 16-bit unsigned integer (in
big-endian encoding) indicating its size.

\item {} 
Further zero bytes are appended up to \code{size\_bl} bytes.

\item {} 
If the payload is already larger than this, then we instead pad up to the
next power-of-two multiple of \code{size\_bl} (e.g. \code{2*size\_bl}, \code{4*size\_bl},
\code{8*size\_bl}, etc) that can contain the payload.

\end{itemize}

\textbf{Trial decryption}. In general, at some points in time, it may be possible to
receive a message from several different sessions. The \code{SIDKEY\_HINT} record
helps to efficiently determine the correct decryption key. We calculate it as
\(H(\mathsf{sid} || \mathsf{gk})[0]\).

Only the first byte of the hash value is used, so it is theoretically possible
that values may collide and subsequent trials may be required to determine the
correct keys to use. However, this should give a performance improvement in the
majority of cases without sacrificing security.

When we implement forward-secrecy ratchets, or when we move to transports that
don't provide explicit hints on the authorship of packets, this technique may
be adapted further to select between the multiple decryption or authentication
keys (respectively) that are available to verify-decrypt packets. Composing
this with anonymity systems will need extra care however - if the hint values
are the same for related messages, then attackers can identify messages sharing
this relationship with greater probability. (This is the case with our scheme
above, where we use the same value for all messages in a subsession.)

\section{Encoding and primitives}
\label{crypto:encoding-and-primitives}
To encode our records, we use TLV (type, length, value) strings, similar to the
OTR protocol. Messages are prepended with the fixed string \code{?mpENC:} followed
by a base-64 encoded string of all concatenated TLV records, followed by \code{.}.
Type and length in the TLV records are each two octets in big-endian encoding,
whereas the value is exactly as many octets as indicated by the length.

This encoding may change at a later date.

The cryptographic primitives that we use are:
\begin{itemize}
\item {} 
Identity signature keys: \href{http://ed25519.cr.yp.to/}{ed25519}

\item {} 
Session exchange keys: \href{http://cr.yp.to/ecdh.html}{x25519}

\item {} 
Session signature keys: ed25519

\item {} 
Message encryption: \href{https://en.wikipedia.org/wiki/Advanced\_Encryption\_Standard}{AES128} in \href{https://en.wikipedia.org/wiki/Block\_cipher\_mode\_of\_operation\#Counter\_.28CTR.29}{CTR mode}

\item {} 
Message references, packet references, trial decryption hints: \href{https://en.wikipedia.org/wiki/SHA-2}{SHA256}

\item {} 
Key derivation: \href{https://tools.ietf.org/html/rfc5869}{HKDF}-SHA256

\end{itemize}

These have generally been chosen to match a general security level of 128 bit
of entropy on a symmetric block cipher. This is roughly equivalent to 256 bit
key strength on elliptic curve public-key ciphers and 3072 bit key strength on
traditional discrete logarithm problem based public-key ciphers (e.g. DSA).

We may switch to a more modern block cipher at a future date, that is more
easily implementable and verifiable in constant time.

\chapter{Future work}
\label{future-work:future-work}\label{future-work::doc}\label{future-work:hkdf}
Here, we discuss tasks for the future, adding functionality and building on our
existing security properties.

\section{Immediate}
\label{future-work:immediate}

\subsection{Missing properties}
\label{future-work:missing-properties}
Some of the security mechanisms that we have described, have not actually been
implemented yet. We list them below. We have delayed them because we consider
them to be less-critical security properties, and wanted to focus on making our
implementation work reliably enough to start testing. However, we recognise
their importance at providing end-to-end guarantees; none of them are expected
to be hard to achieve, and solution outlines have already been sketched out.
\begin{description}
\item[{Error and abort messages}] \leavevmode
We should add the ability to abort a membership change, either manually or
automatically after a timeout. Currently if someone disconnects, others will
be left waiting until they themselves leave the channel. This must be done
via an explicit ``abort'' packet that gets handled by the concurrency resolver,
to prevent races between members.

For liveness properties, we already have timeouts that emit local security
warnings if good conditions are not reached after a certain time. However,
user experience will benefit if we have explicit error messages that inform
other members of conditions such as ``inconsistent history'', that \emph{fail-fast}
more quickly than our generous default timeouts.

\item[{Server-order consistency checks}] \leavevmode
We need to implement consistency checks for accepted operations, since we use
the server's packet ordering. This is similar to our consistency checks for
messages: expect everyone to send authenticated acks to confirm their view of
all previous operations. Further, the parent reference contained within the
initial packet must also be authenticated by the initiator.

Together, these prevent the server from re-ordering operations \emph{completely}
arbitrarily. However, given a set of concurrent operations with the same
authenticated parent, it is still able to choose which one is accepted, by
broadcasting that one first. Given other constraints, we think this security
tradeoff is acceptable. \footnote[1]{\sphinxAtStartFootnote%
Users may manually retry rejected operations as many times as they
want, and it would be extremely suspicious if it is rejected too often. Note
that automatic retries are a security risk.
}

\item[{Flow control}] \leavevmode
We need recovery (automatic resends) for dropped messages, and heartbeats to
verify in-session freshness. (We already have security warnings for messages
received out-of-order.) This is already implemented in a non-production-ready
Python prototype, with random integration tests, and only needs to be ported.

\end{description}
\phantomsection\label{future-work:publish-sess-sig-keys}\begin{description}
\item[{Publish signature keys}] \leavevmode
We need to make signature-key publishing logic work properly, so that we have
deniable authentication for message content.

Roughly, once a member makes a subsession shutdown request (\code{FIN}), they
may publish their signature key after everyone acks this request. This is
safe (i.e. an attacker cannot reuse the key to forge messages), if we enforce
that one may not author messages \emph{after} a \code{FIN}, i.e. all receivers must
refuse to accept such messages. However, this simple approach destroys our
ability to authenticate our own acks of others' messages (e.g. \emph{their}
\code{FIN}) after we send our own \code{FIN}. So we'll need something a bit more
complex, and we haven't worked out the details yet.

If others' acks to our \code{FIN} are blocked, then we will never be sure that
it's safe to publish our signature key. This likely can't be defended under
this type of scheme, since confidential authenticity isn't meaningful without
authenticity (it would be ``confidential nothing''); the equivalent attack also
applies to OTR. To defend against this, we would need a session establishment
protocol that is itself deniable, and then we don't need to mess around with
publishing keys; see ``Better membership change protocol'' below.

\end{description}

\section{Next steps}
\label{future-work:next-steps}

\subsection{Security improvements}
\label{future-work:security-improvements}\begin{description}
\item[{Messaging ratchet for intra-subsession forward secrecy}] \leavevmode
We already have forward secrecy for old subsessions, but this is important
for long-running subsessions and later when we do asynchronous messaging.

One simple scheme is to deterministically split the key into \(n\) keys,
one for each sender. Then, each key can be used to seed a hash-chain ratchet
for its associated sender. Once all readers have decrypted a packet and
deleted the key, the forward secrecy of messages encrypted with that key and
previous ones is ensured. However, since this scheme does not distribute
entropy between members, there is no chance to recover from a memory leak and
try to regain secrecy for future messages.

\item[{Better membership change protocol}] \leavevmode
Use a constant-round group key exchange such as that from \phantomsection\label{future-work:id2}{\hyperref[references:np1sec]{\crossref{{[}np1sec{]}}}}, or even
pairwise \phantomsection\label{future-work:id3}{\hyperref[references:triple\string-dh]{\crossref{{[}Triple-DH{]}}}} between all group members which extends better to
asynchronous messaging. In both cases, we get deniability for free without
having to publish signature keys for messages.

\item[{Use peer-to-peer or anonymous transport for non-GKA messages}] \leavevmode
The only part of our system that requires a linear ordering is the membership
operation packets. So it is possible to move message packets into a transport
that gives us better properties (e.g. against metadata analysis) than XMPP.

\end{description}

\subsection{More functionality}
\label{future-work:more-functionality}\begin{description}
\item[{Large messages and file transfer}] \leavevmode
Our current padding scheme limits messages to roughly \(2^{16}\) bytes,
to keep it under our XMPP server maximum stanza size. This may be extended to
arbitrary sizes: pad to the next-power-of-2-multiple of the maximum size, and
split this into MTU-sized packets. (This is just a functionality improvement;
these padding schemes have not been researched from an adversarial model.)
After this, it is a straightforward engineering task to allow file transfers.

\item[{Better multi-device support}] \leavevmode
A group messaging system can already be used for multiple devices in a very
basic way -- simply have each device join the session as a separate member.
This is desireable for security reasons, since it means we can avoid sharing
ephemeral keys. It's unclear whether devices should share identity keys, or
use different identity keys and have the PKI layer bind them to say ``we are
the same identity'', but this decision doesn't affect our messaging system.

Beyond this, we can add some things both in the messaging layer and in the UI
layer to make the experience smoother for users.
\begin{itemize}
\item {} 
The users view should show one entry per user, not per device;

\item {} 
Not-fully-acked warnings may be tweaked to only require one ack from every
\emph{user} rather than every device. However, a warning should probably fired
eventually even if all devices don't ack it, just later than in the single
device case; it is still a critical security error if different devices get
\emph{different} content. Similar logic may be applied to heartbeats;

\item {} 
There are extra corner cases in the browser case, where the user may open
several tabs (each acting as a separate device), with crashes and page
reloads causing churn that might reveal implementation bugs.

\end{itemize}

(We are already doing some of the above.)

\item[{Sync old session history across devices}] \leavevmode
It is unnecessary to reuse security credentials (e.g. shared group keys or
session keys) that are linked to others -- we already decrypted the packets
and don't need to do this again. Futher, credentials in modern protocols are
supposed to be ephemeral, and this is a vital part of their security. If we
retain such credentials, we may put others at risk or leave forensic traces
of our own activities.

Therefore, our sync mechanism must not directly reuse ciphertext from our
messaging protocol, since it forces us to store these credentials. It is much
better to re-encrypt the plaintext under our own keys, unlinked to anyone
else. That is, \emph{at the very least}, this feature must be a separate protocol;
the security model here is \emph{private storage} for oneself, and \emph{not} private
communications. Finally, even following this requirement, long-term storage
of encrypted data directly counteracts forward secrecy, so the user must be
made aware of this before such a feature is enabled.

\end{description}

\section{Research}
\label{future-work:research}
Here are some research topics for the future for which we have no concrete
solution proposals, though we do have some vague suggestions.

Several of these relate to ``no-compromise'' asynchronous messaging, i.e. with
causal ordering, no breaking of symmetry between members, no requirement of
temporary synchronity or total ordering, no accept-reject mechanisms, and no
dependency on external infrastructure.
\begin{description}
\item[{Merging under partial visibility}] \leavevmode
As mentioned earlier, our membership operations are in a total order because
nobody has defined how to merge two group key agreements. This problem has a
well-defined solution for pairwise key agreements, but only if everyone can
see all history, or if only member inclusions are allowed (or generally, if
the operations to be merged have no inverse). If we have partial visibility
(i.e. members can't see events from before they join) \emph{and} we want to
support both member inclusion and exclusion, the solution is unknown.

\item[{Session rejoin semantics}] \leavevmode
As part of solving the above point, we need to decide what parent references
mean exactly in the context of rejoining a session. Existing members' parent
references to older messages won't make sense to us since we can't see them;
symmetrically, we might want to reference the last messages we saw before
previously leaving the session, but these references might not make sense to
some of the existing members, i.e. those not present when we parted.

\item[{Possible hybrid solution}] \leavevmode
One possible solution is to allow causally-ordered member inclusion, but
require that everyone acknowledge a member exclusion before it is considered
complete. Then our partial visibility problem disappears; new members don't
have to worry about how to merge in excludes that happened before they joined
-- their inviter will have already taken this into account. This is probably
the least non-zero ``compromise'' solution, but the agreement mechanism might
itself be very complex.

\item[{Save and load current session}] \leavevmode
This is vital for asynchronous messaging, and would be a straightforward but
significant engineering effort on top of our existing implementation.

One optimisation to be made after the basic ability is complete, is to prune
older messages from our transcript and message-log data structures. This must
be thought through carefully, since we need a limited set of history in order
to perform ratcheting, check the full-ack status of messages and freshness of
other members, and merge concurrent membership operations.

\item[{Membership change \emph{policy} protocol}] \leavevmode
This ought to be layered on top of a membership change \emph{mechanism} protocol.
When reasoning about security, naturally one considers who is allowed to do
what. But authorization is a separate issue from \emph{how to execute membership
changes}. We should solve the latter first, assuming that all members are
allowed to make any change (in many cases this is exactly what is desired),
\emph{then} think about how to construct a secure mechanism to restrict these
operations based on some user-defined policy. This is the same reason why we
generally perform authentication before, and separately from, authorization.

\end{description}

\renewcommand{\indexname}{Index}
\printindex
\end{document}

%% file: unicode.tex
\usepackage{amssymb}
\DeclareUnicodeCharacter{21A6}{$\mapsto$}
\DeclareUnicodeCharacter{2190}{$\leftarrow$}
\DeclareUnicodeCharacter{2194}{$\leftrightarrow$}
\DeclareUnicodeCharacter{21D2}{$\Rightarrow$}
\DeclareUnicodeCharacter{21D4}{$\Leftrightarrow$}
\DeclareUnicodeCharacter{2200}{$\forall$}
\DeclareUnicodeCharacter{2203}{$\exists$}
\DeclareUnicodeCharacter{2204}{$\nexists$}
\DeclareUnicodeCharacter{2205}{$\emptyset$}
\DeclareUnicodeCharacter{2208}{$\in$}
\DeclareUnicodeCharacter{2209}{$\notin$}
\DeclareUnicodeCharacter{2227}{$\wedge$}
\DeclareUnicodeCharacter{2228}{$\vee$}
\DeclareUnicodeCharacter{2229}{$\cap$}
\DeclareUnicodeCharacter{2260}{$\neq$}
\DeclareUnicodeCharacter{2261}{$\equiv$}
\DeclareUnicodeCharacter{2264}{$\leq$}
\DeclareUnicodeCharacter{2265}{$\geq$}
\DeclareUnicodeCharacter{2282}{$\subset$}
\DeclareUnicodeCharacter{2286}{$\subseteq$}
\DeclareUnicodeCharacter{2288}{$\nsubseteq$}
\DeclareUnicodeCharacter{228F}{$\sqsubset$}
\DeclareUnicodeCharacter{22A5}{$\perp$}